\documentclass[a4paper,prb,onecolumn,superscriptaddress]{revtex4-2}
\usepackage[dvipsnames]{xcolor}
\usepackage{amsmath}
\usepackage{amsfonts}
\usepackage{graphicx}
\usepackage{amssymb}
\usepackage{siunitx} 
\usepackage{multirow}

\usepackage[normalem]{ulem}

\newcommand{\colvec}[2][.8]{%
  \scalebox{#1}{%
    \renewcommand{\arraystretch}{.8}%
    $\begin{pmatrix}#2\end{pmatrix}$%
  }
}

\newcommand{\avec}{\mathbf{a}}

\newcommand{\nvec}{\mathbf{n}}
\newcommand{\qvec}{\mathbf{q}}
\newcommand{\kvec}{\mathbf{k}}
\newcommand{\Pvec}{\mathbf{P}}

\begin{document}

\title[Superlight pairs in face-centred-cubic extended Hubbard models with strong Coulomb repulsion]{Superlight pairs in face-centred-cubic extended Hubbard models with strong Coulomb repulsion}

\author{G D Adebanjo}
\email{ganiyu.adebanjo@gmail.com}

\affiliation{School of Physical Sciences, The Open University, Walton Hall, Milton Keynes, MK7 6AA, UK}

\author{P E Kornilovitch}
\affiliation{Department of Physics, Oregon State University, Corvallis, OR, 97331, USA}

\author{J P Hague}
\email{jim.hague@open.ac.uk}
\affiliation{School of Physical Sciences, The Open University, Walton Hall, Milton Keynes, MK7 6AA, UK}

\date{\today}

\begin{abstract}

The majority of fulleride superconductors with unusually high transition-temperature to kinetic-energy ratios have a face-centred-cubic (FCC) structure.
We demonstrate that, within extended Hubbard models with strong Coulomb repulsion, paired fermions in FCC lattices have qualitatively different properties than pairs in other three-dimensional cubic lattices.
Our results show that strongly bound, light, and small pairs can be generated in FCC lattices across a wide range of the parameter space. We estimate that such pairs can Bose condense at high temperatures even if the lattice constant is large (as in the fullerides). 
\end{abstract}
\maketitle

\section{Introduction}

Superlight small pairs are of interest in the context of superconductivity due to their potential to form Bose-Einstein condensates (BEC) at high temperatures \cite{hague2007superlighta,hague2007superlightb}. There are a number of low-dimensional systems within which superlight pair states can be realised, for example the staggered ladder \cite{hague2007superlightb}, triangular lattice \cite{hague2007superlighta,hague2008_sing_trip_bip_triangular} and quasi-two-dimensional hexagonal lattice \cite{hague2010Lightandstable}. Pairs consisting of two fermions can be bound onto neighbouring sites by a combination of strong intersite attraction and strong onsite repulsion. Such pairs can be light and small (superlight) if it is possible to move to neighbouring lattice sites via a single hop without breaking the pairing  \cite{hague2007superlightb}, so that the pair motion is a first order effect. 
In many materials there is a strong onsite Coulomb repulsion, so intersite pairs are formed via any intersite or long-range attraction, which could originate either from phonons or other more exotic mechanisms. 
The aim of this article is to explore the possibility of superlight small pairs in FCC lattices.

Extended Hubbard models \cite{hirsch1984,micnas1990superconductivity} contain the essential interactions to realise superlight states. The Hamiltonian of an extended Hubbard model is defined as:
\begin{equation}\label{themodelHamiltonian}
H = \sum_{\langle\nvec,\avec\rangle\sigma}t_{\avec}\,c_{\nvec+\avec,\sigma}^{\dagger}\,c_{\nvec\sigma} + U\sum_{\nvec} \hat{\rho}_{\nvec \uparrow}\,\hat{\rho}_{\nvec \downarrow} +   \sum_{\langle\nvec,\avec\rangle}V\:\hat{\rho}_{\nvec + \avec}\:\hat{\rho}_{\nvec}
\end{equation}
where $c^{\dagger}_{\nvec\sigma}$ ($c_{\nvec\sigma}$) creates (annihilates) an electron of spin $\sigma$ at site $\nvec$, $\hat{\rho}_{\nvec}=\hat{\rho}_{\nvec\uparrow}+\hat{\rho}_{\nvec\downarrow}$, where $\hat{\rho}_{\nvec\sigma}$ is the number operator for electrons on site $\nvec$ with spin $\sigma$, $\avec$ is the intersite lattice vector, $t_{\avec}$ is the intersite hopping, $U$ is the onsite interaction and $V$ is the intersite interaction. Both $U$ and $V$ may be attractive or repulsive, although in most materials repulsive $U$ is more likely due to the difficulties of overcoming the Hubbard $U$ with attractive interactions, such as those due to electron-phonon interactions. In the low-density limit the model is also known as the $UV$ model.
Properties of local pairs, which can be used to estimate the Bose-Einstein condensation temperature, have been studied in simple systems using the $UV$ model \cite{kornilovitch2004,hague2010Lightandstable,Mbak2007,Davenport2012,adebanjo2021fermion}. If $U$ is highly repulsive and $V$ is attractive, then superlight pairs can be found on suitable lattices. 

Extended Hubbard models have been extensively applied to the quasi-2D cuprate superconductors \cite{Micnas_1988,PhysRevB.97.184507}. The origin of an intersite $V$ can be from Coulomb repulsion, long range electron-phonon interactions \cite{PhysRevB.79.212501}, and an intersite $J$ can originate from anti-ferromagnetic interactions induced by the Hubbard $U$ \cite{jozef2017}. Various phases are predicted in extended Hubbard models, such as spin triplet pairing \cite{qu}, $d$-wave superconductivity \cite{PhysRevB.97.184507}, Mott insulators \cite{hubbard1963electron}, XY antiferromagnetism \cite{Laad_1991} and stripe order \cite{kato2000}. We note that  experimental evidence for strong intersite attractions mediated by phonons has been reported recently in one-dimensional cuprates \cite{chen2021a,wang_chen_et_al_2021}. 

\begin{figure*}
	\centering
	\includegraphics[width=175mm]{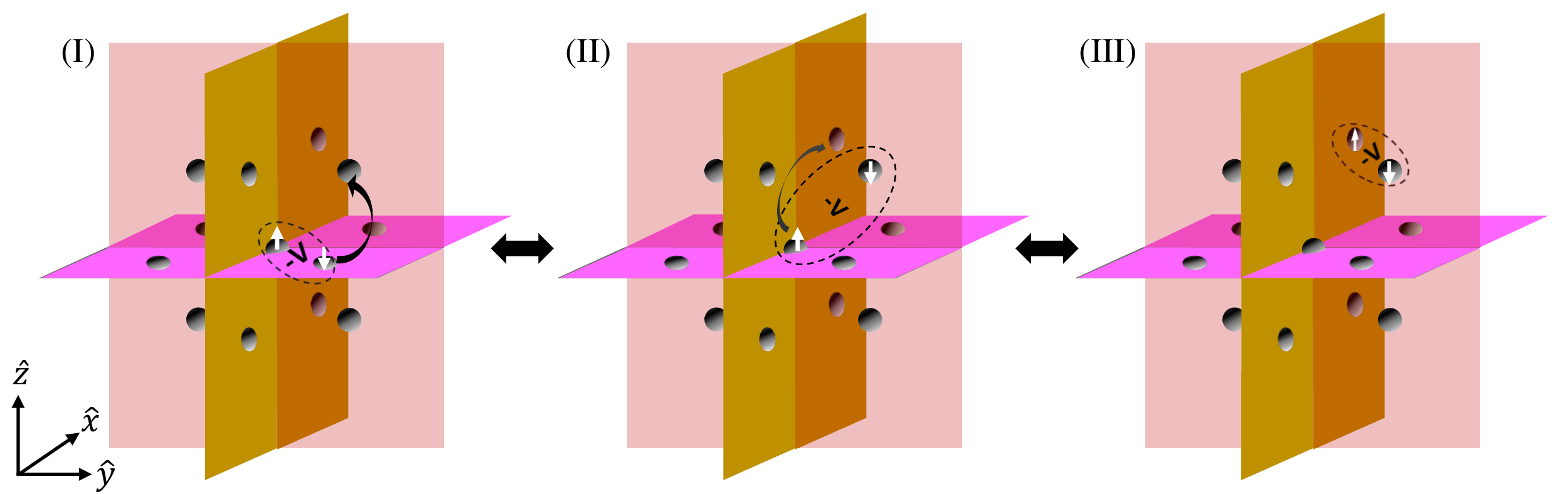}
	\caption{[Color online] Schematic demonstrating the first-order hopping of superlight small pairs on the FCC lattice. The white vertical arrow represents an electron with its spin, each small gray circle is a lattice site (only the 13 sites of interest are shown), the dashed oval represents a bound state through an attractive $V$, the curved arrow represents electron hopping, and the two-way arrow implies that the configurations are degenerate. The key feature here is that the pair is itinerant as long as the intersite interaction is sufficiently attractive, and there is sufficient Hubbard $U$ to stop on-site pairs from forming. Unlike in other cubic lattices where first-order superlight states are only attainable via an attractive $U$, superlight state in the FCC lattice corresponds to a more physical case where the local Coulomb repulsion is large.}
	\label{superlight_hopping_fcc}
\end{figure*}

In face-centred cubic (FCC) lattices, electrons paired between near-neighbour sites can move with a single hop. An illustration of such pair movement in an FCC lattice is shown in Fig. \ref{superlight_hopping_fcc}. If sufficient intersite attraction is present, and there is repulsive Hubbard $U$ to suppress on-site pairing, the pair can move easily through the lattice. This should result in a low effective pair mass for small pairs, which could in turn yield a high transition temperature.  To our knowledge, superlight pairs have not yet been examined in FCC systems.  The complexity of the FCC lattice structure and increased number of nearest-neighbour sites complicate the calculation and we aim to fill this gap. The detailed calculations presented in this paper explore how pair properties evolve with Hubbard $U$ and $V$ in FCC lattices, and identify regions of the parameter space where pairs are small and light.

The microscopic understanding of complex phases in strongly correlated systems remains a challenge. There are many numerical techniques that are suitable for treating strongly correlated systems, including extended Hubbard models. These include exact diagonalisation (ED) \cite{ARRACHEA2004224}, density matrix renormalisation group (DMRG) and matrix product state calculations \cite{SCHOLLWOCK201196,PhysRevB.76.155121}, dynamical mean-field theory (DMFT), its extensions dynamical cluster approximation (DCA), cellular DMFT (CDMFT), and extended DMFT to treat extended Hubbard models \cite{RevModPhys.68.13,RevModPhys.77.1027,PhysRevB.96.235149,PhysRevB.99.205156,PhysRevB.97.184507}, quantum Monte Carlo (QMC) techniques \cite{hague2007superlighta}, and recently quantum embedding \cite{lupo,lee2019,albadri2020} and machine learning algorithms \cite{PhysRevB.104.205120}. Many of these techniques are limited to 1D and 2D systems, either inherently, or because particle numbers are limited. The exponentially growing Hilbert space limits ED to small numbers of sites and particles, effectively limiting application to 1D and small 2D systems. DMRG and MPS work best in 1D. QMC techniques can suffer from sign problems when the number of particles becomes large, although the number of sites may not be limited. On the other hand, DMFT is most accurate for large spatial dimensions, although the coarse graining of the self-energy removes some details of the lattice \cite{RevModPhys.68.13}. The integration of DMFT and density functional theory calculations has led to powerful techniques for the simulation of materials \cite{linscott2020}.

In spite of their ubiquity in condensed matter systems, FCC lattices are often overlooked within the correlated electrons community owing to their relative complexity compared to other lattices.  Materials of interest with FCC lattices include the A$_3$C$_{60}$ compounds: a family of molecular compounds with high transition temperature \cite{gunnarsson1997superconductivity} (where A is an alkali metal e.g. K, Rb, Cs) which are predominantly FCC structured \cite{gunnarsson2004alkali}. In addition to electron-phonon interactions \cite{gunnarsson1997superconductivity} found in these alkali-doped compounds, strong correlation  \cite{capone2009colloquium} is also prevalent. The presence of long-range phonon mediated interactions (e.g. the intermolecular modes \cite{gunnarsson2004alkali,takada1998superconductivity}) may lead to suitable conditions for extended Hubbard physics and superlight pairs could also be relevant to other FCC materials.

This work aims to provide an exact solution of the two-electron problem in an FCC lattice. We calculate the critical potentials $U_c$ $(V_c)$ to bind particles into pairs, the system's total energy, the pair's size and mass, and BEC transition temperatures of pairs in the low-density (dilute) limit. The paper is organised as follows: We describe the model Hamiltonian and methodology used to solve the $UV$ model in the dilute limit (Sec. \ref{sec:method_uv_on_fcc}). In Sec. \ref{sec:results}, the properties of the formed pairs are reported. We conclude this work with a discussion in Sec. \ref{sec:discussion}.
\section{Methodology}
\label{sec:method_uv_on_fcc}

We find exact solution to a single-orbital system of two spin 1/2 fermions where the orbital energy is taken to be zero.  The two-body problem relevant to the Hamiltonian in Eqn. (\ref{themodelHamiltonian}) must satisfy the equation below:

\begin{align}
& \sum_{\avec}t_{\avec}\,\bigg[\Psi(\nvec_{1}+\avec,\nvec_{2}) +\Psi(\nvec_{1},\nvec_{2}+\avec)\bigg] + \sum_{\avec}\hat{V}_{\avec}\,\delta_{\nvec_{1} - \nvec_{2},\avec}\Psi(\nvec_{1},\nvec_{2}) = E\,\Psi(\nvec_{1},\nvec_{2}) 
\label{twobody wavefunction S.E}
\end{align}
\noindent Here $E$ is the total energy of the system, $\Psi(\nvec_{1},\nvec_{2})$ is the real-space wave function of the fermions, and $\nvec_{1}$ and $\nvec_{2}$ are the spatial coordinates.  $\hat{V}_{\avec}$ combines the interaction terms into a single function with $\hat{V}_{r=0}=U$ and $\hat{V}_{r=\avec }=V$. The wave function in Fourier space is
\begin{equation}\label{FT wave function}
\psi_{\kvec_{1}\kvec_{2}} = \frac{1}{N}\sum_{\nvec_{1}\nvec_{2}}\Psi(\nvec_{1},\nvec_{2})\,e^{-i\kvec_{1}\,\nvec_{1} - i\kvec_{2}\,\nvec_{2}}
\end{equation}
\noindent where $N$ is the total number of lattice points. Substituting the inverse of Eqn. (\ref{FT wave function}) into Eqn. (\ref{twobody wavefunction S.E}) gives:
\begin{equation}\label{equation_four}
(E-\varepsilon_{\kvec_{1}}-\varepsilon_{\kvec_{2}})\psi_{\kvec_{1}\kvec_{2}}=\frac{1}{N}\sum_{\avec\qvec}\hat{V}_{\avec}\,e^{i(\qvec -\kvec_{1})\avec}\,\psi_{\qvec,\kvec_{1}+\kvec_{2}-\qvec} \;.
\end{equation}
where  $\kvec$ is the particle's momentum vector.
The dispersion relation in the FCC lattice is given as
\begin{align}\label{one_p_disp_relation} 
\begin{split}
\varepsilon_{\kvec} & = -4t\bigg[\cos\frac{k_xb}{2}\cdot\cos\frac{k_yb}{2} + \cos\frac{k_yb}{2}\cdot\cos\frac{k_zb}{2} + \cos\frac{k_xb}{2}\cdot\cos\frac{k_zb}{2} \bigg] 
\end{split}
\end{align}
\noindent where $b$ is the lattice constant.

The solution to the problem involves 13 self-consistent algebraic equations. We apply (anti-)symmetrisation to separate the symmetric (singlet) states from the anti-symmetric (triplet) states. Following Ref. \cite{adebanjo2021fermion}, the (anti-)symmetrised wave function is expressed as
\begin{align}\label{equation_thirteen}
& (E-\varepsilon_{\kvec_{1}}-\varepsilon_{\kvec_{2}})\phi_{\kvec_{1}\kvec_{2}}^{\pm} = \frac{1}{N} \sideset{}{'}\sum_{\qvec\avec}\hat{V}_{\avec}\;\Big\{e^{i(\qvec -\kvec_{1})\,\avec} \pm e^{i(\qvec -\kvec_{2})\,\avec} \Big\}\;\phi^{\pm}_{\qvec,\kvec_{1}+\kvec_{2}-\qvec} 
\end{align} 
where 
\begin{equation}\label{equation_twelve}
\phi_{\kvec_{1}\kvec_{2}}^{\pm}=\psi_{\kvec_{1}\kvec_{2}} \pm \psi_{\kvec_{2}\kvec_{1}} \;\;\;\;\; ,
\end{equation}
 and $+$ and $-$ refer to the singlet and the triplet wave functions, respectively. 
 The summation over the lattice vector, $\avec$, in Eqn. (\ref{equation_thirteen}) is split into two sets: $\{\avec_+\}$ for singlets, and $\{\avec_-\}$ for triplets. We define them as
\begin{align}
    \begin{split}
    \{\avec_{+}\} & = \{ (0,0,0), (\frac{b}{2},\frac{b}{2},0), (0,\frac{b}{2},\frac{b}{2}), (\frac{b}{2},0,\frac{b}{2}), (\frac{b}{2},-\frac{b}{2},0),(0,\frac{b}{2},-\frac{b}{2}), (-\frac{b}{2},0,\frac{b}{2}) \}
    \end{split}
    \\
\{\avec_{-}\} 
& = \{ (\frac{b}{2},\frac{b}{2},0), (0,\frac{b}{2},\frac{b}{2}), (\frac{b}{2},0,\frac{b}{2}), (\frac{b}{2},-\frac{b}{2},0), (0,\frac{b}{2},-\frac{b}{2}), (-\frac{b}{2},0,\frac{b}{2})\} 
\label{schrodinger_equation_trip}
\end{align}
where, again, $b$ is the lattice constant. The primed summation in Eqn. (\ref{equation_thirteen}) means that a factor of 1/2 should be included for the case $\avec_{+}=0$.
If we define a function 
\begin{equation}\label{equation_six}
\Phi_{\avec_{\pm}}^{\pm}(\kvec_{1}+\kvec_{2})=\Phi_{\avec_{\pm}}^{\pm}(\Pvec)\equiv \frac{1}{N}\sum_{\qvec}e^{i\qvec\avec_{\pm}}\;\phi_{\qvec,\Pvec-\qvec}^{\pm}
\end{equation}
where $\Pvec=\kvec_{1}+\kvec_{2}$ is the total momentum of the particle pair, then, $\phi_{\kvec_{1}\kvec_{2}}^{\pm}$ in Eqn. (\ref{equation_thirteen}) can be expressed in terms of $\Phi_{\avec_{\pm}}^{\pm}$. Replacing $\phi_{\qvec,\Pvec-\qvec}^{\pm}$ in Eqn. (\ref{equation_six}) leads to the self-consistent equations
\begin{equation}\label{equation_seventeen}
\Phi_{\avec_{\pm}}^{\pm}(\Pvec)=- \sum_{\avec_{\pm}^{'}}\hat{V}_{\avec_{\pm}}
L_{\avec_{\pm}\,\avec_{\pm}^{'}}^{\pm}(E,\Pvec)\;\Phi_{\avec_{\pm}^{'}}^{\pm}(\Pvec)
\end{equation}
where
\begin{equation}\label{equation_eighteen}
L_{\avec_{\pm}\,\avec_{\pm}^{'}}^{\pm}(E,\Pvec)=\frac{1}{N}\sum_{\qvec}\frac{e^{i\qvec(\avec_{\pm}-\avec_{\pm}^{'})}\pm e^{i[\qvec\avec_{\pm}-(\Pvec-\qvec)\avec_{\pm}^{'}]}}{-E+\varepsilon_{\qvec}+\varepsilon_{\Pvec-\qvec}}
\end{equation}
is the Green's function of the lattice. The full dispersion matrices for the singlets and triplets can be found in Appendix \ref{appendix:uv_derivation} and all calculations are at zero temperature. In the thermodynamic limit (infinite lattice size), Equation (\ref{equation_eighteen}) is a generalised Watson integral that in 3D converges for any energy that’s below the threshold energy of $-2W$.

\section{Results} \label{sec:results}

\begin{figure}[h!]
    \centering
	\includegraphics[scale=0.5]{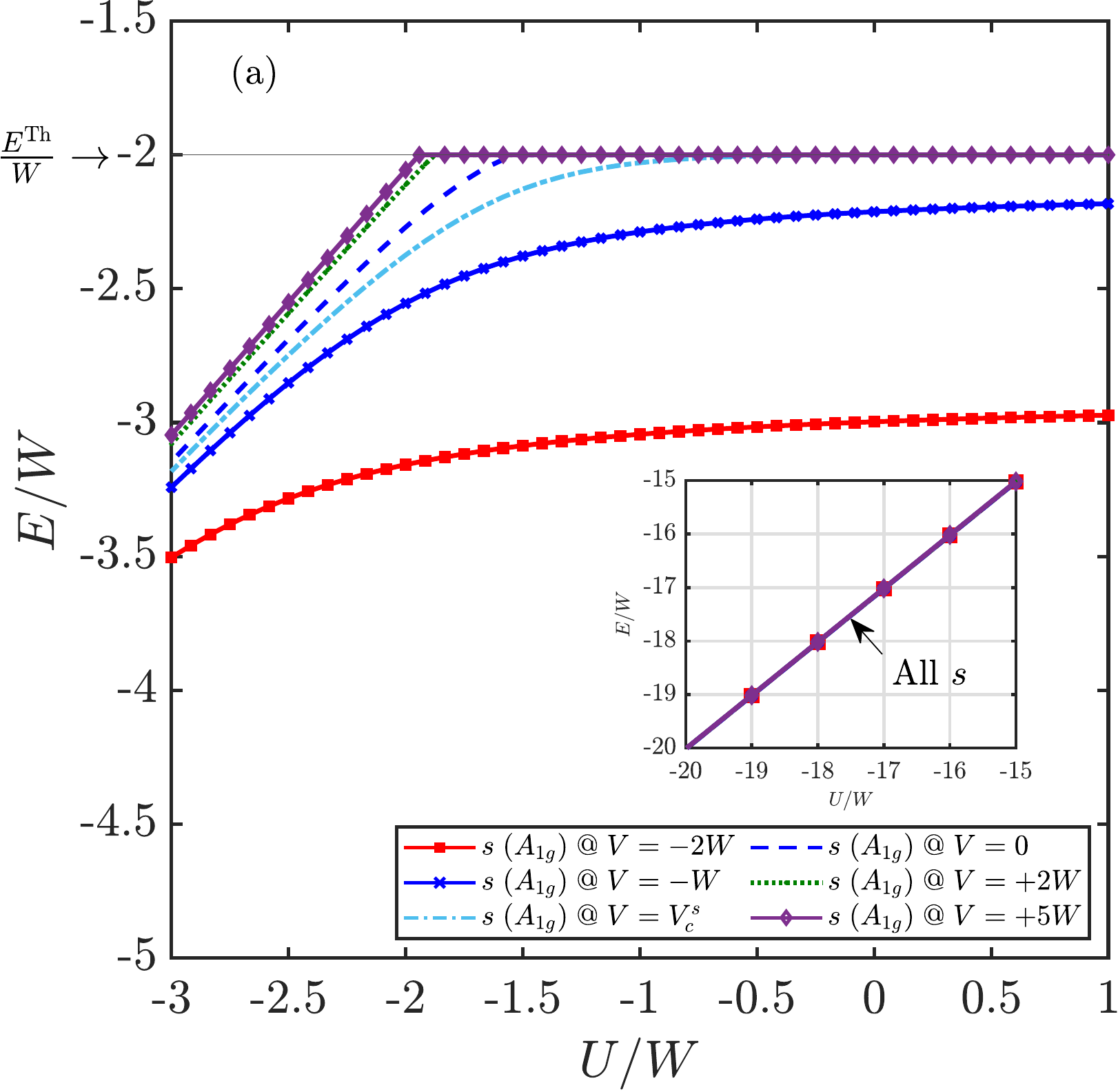}
	\vspace{0.8em}
	\includegraphics[scale=0.5]{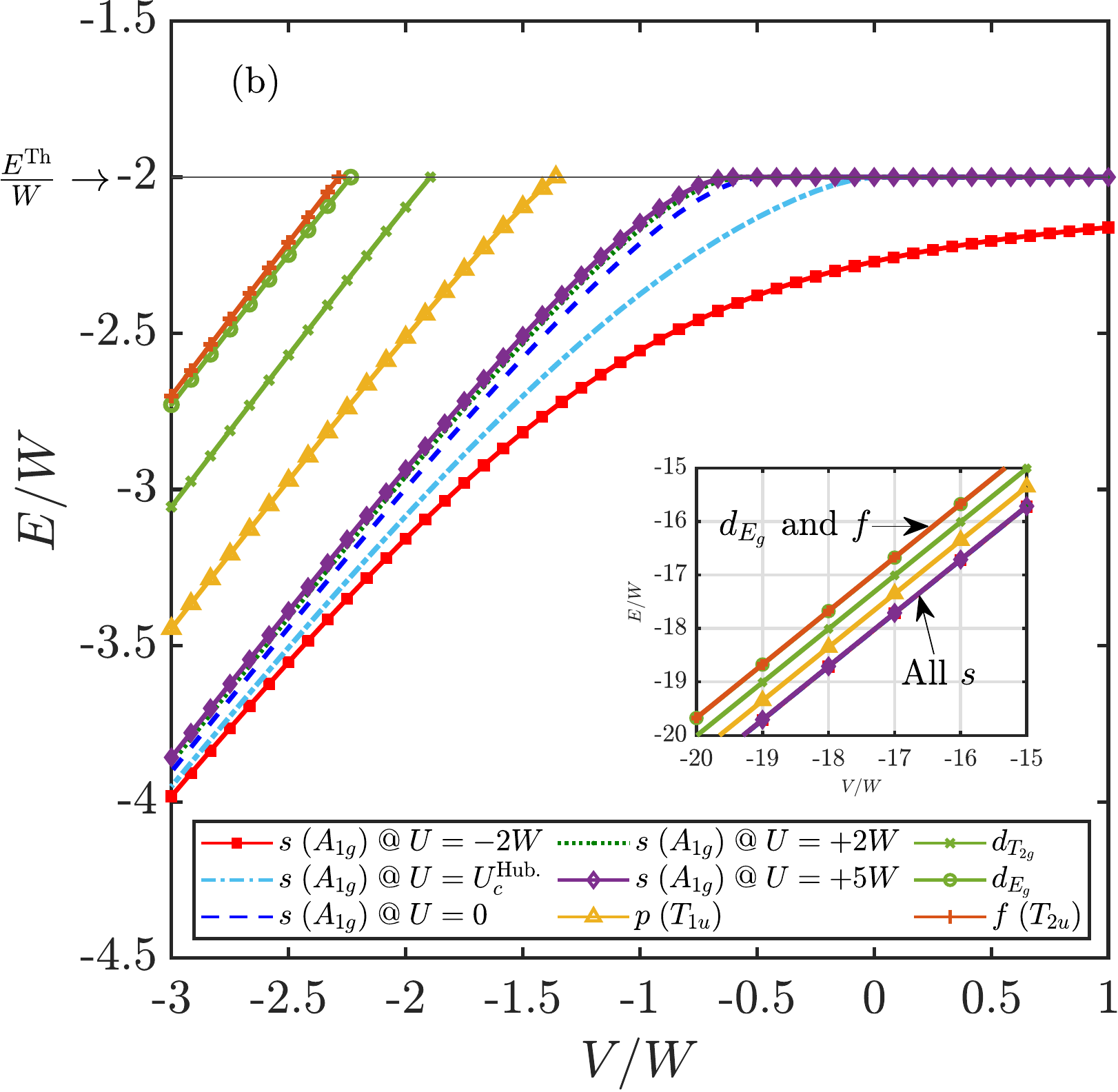}
	\caption{The total energy of pairs for (a) the $s$-states only, and (b) all states of various symmetries. The critical attractions for the $s$-state are $U^{\rm Hub.}_{c}$($V$=0)=$-1.4874W$ and $V^{s}_{c}$($U$=0)=$-0.4836W$. All states except the $s$-state are unaffected by $U$. The excited states ($p$, $d_{T_{2g}}$, $d_{E_{g}}$ and $f$) only appear at strongly attractive $V$. At very large attractions $U$ and $V$, all $s$-states have similar values (insets of panels (a) and (b)). For large intersite attraction, $V\rightarrow-\infty$, the $d_{E_{g}}$ and $f$ states have approximately the same energies and are indiscernible: inset of panel (b).}
	\label{fig:total_energy}
\end{figure}

\subsection{Total Energy}

The ground state energy can be used to identify whether two particles are bound or not. At zero momentum and zero temperature, the threshold energy of two unbound particles is $E^{\rm Th}$=$-2W$, where $W=12t$ is the half-bandwidth. The total energy in Fig. \ref{fig:total_energy} shows all the different pair symmetries found in the FCC lattice. The particles are unbound when there is a plateau at $-2W$ and the energy drops below this threshold value when the particles bind.

A critical attraction must be reached before the formation of any bound state.
At very large attractions $U$ and $V$, the $s$-states have very similar values (insets of Fig. \ref{fig:total_energy}a and \ref{fig:total_energy}b). In Fig. \ref{fig:total_energy}b, the $s$-states are formed at relatively weak intersite attractions $V$ compared to other states.  Additionally, for very strong attractive $V$, the energies of all the pair symmetries are separated by an energy of order $t$: except for the $d_{E_{g}}$- and $f$- states that have approximately the same energies and are therefore indiscernible when plotted (inset of Fig. \ref{fig:total_energy}b). Note that the labels $d_{T_{2g}}$ and $d_{E_{g}}$ are $d$-states with $T_{2g}$ and $E_{g}$ symmetries respectively. The separation of states at large attractive $V$ occurs because the matrix elements only depend on the hopping parameter $t$ as $V\rightarrow -\infty$, as can be seen in Eqn. (\ref{twopart:eq:appgeleven}). Put another way, at deep $V$ the particles are confined to a fixed size shell, and therefore the region of confinement, and thus the energies, become $V$ independent.  At $\Pvec=0$, the degeneracy of the $p$-, $d_{T_{2g}}$-, $d_{E_{g}}$-, and $f$- states is three-fold, three-fold, two-fold and three-fold, respectively. Spin singlet ground states of pairs are common in Hubbard models \cite{byczuk1992}.
\begin{figure}[h!]
	\centering
	\includegraphics[width=98mm]{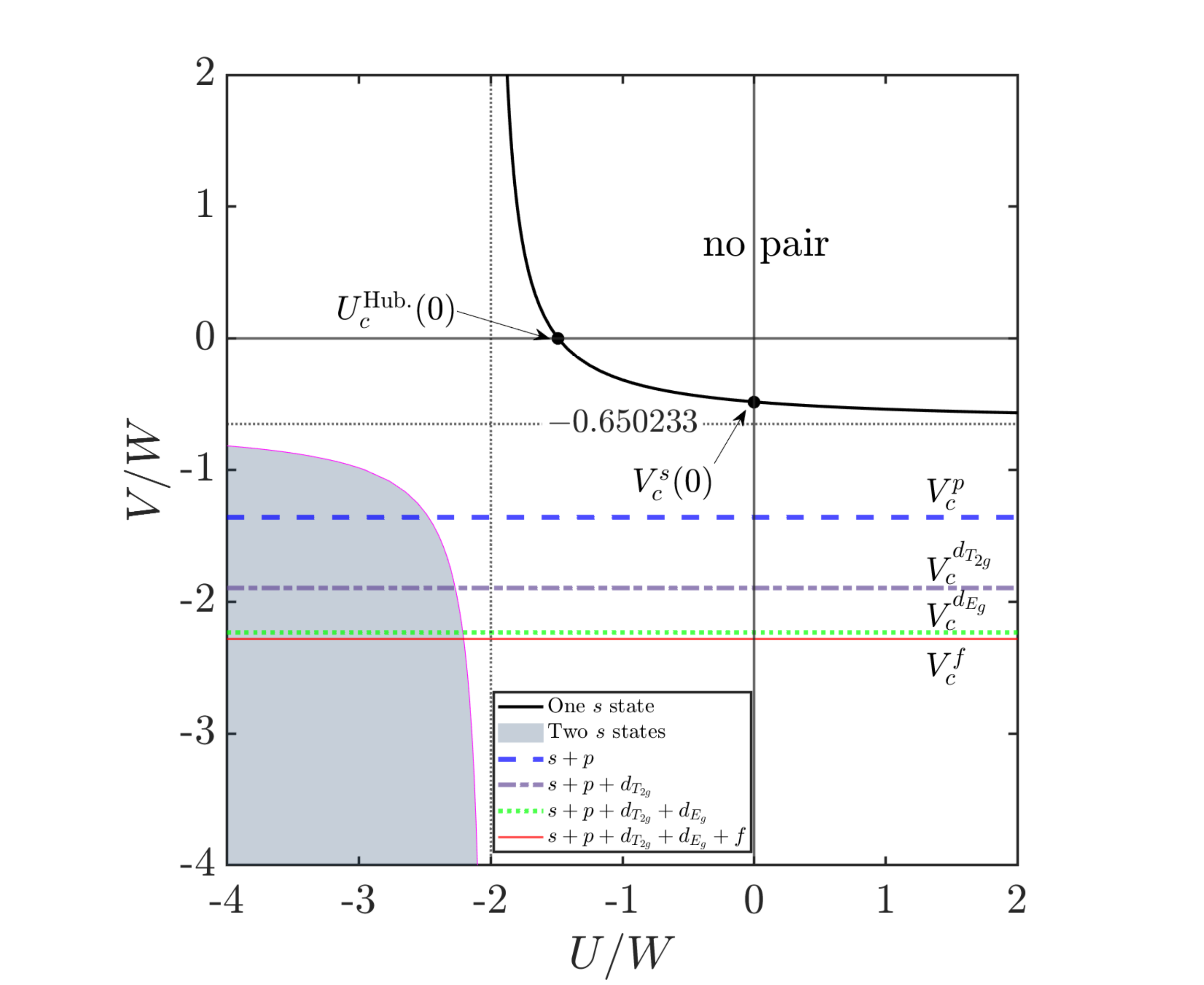}
	\caption{[Color online] Binding diagram showing pair formation at $\Pvec=0$ in an FCC lattice. The top (black) curved line is the boundary that separates bound from unbound $s$-symmetry pairs, the (grey) shaded region enclosed by the (magenta) solid line indicates a region with two $s$-states, the (blue) thick dashed line represents the onset of triply degenerate $p$-states, the (purple) dash-dotted line shows the binding of a triple degenerate $d$-symmetry pair of $T_{2g}$ symmetry (labelled $d_{T_{2g}}$), the (green) dotted line is the line below which two $d$-wave pairs with $E_g$ symmetry start to form (labelled $d_{E_{g}}$) and the (red) solid line indicates the formation of a triply-degenerate $f$ pairs. An $s$-state is guaranteed to form for attractions equal to or stronger than $U=-2W$, $V=-0.65023W$. The boundaries are exact to at least 8 significant figures.}
	\label{fig:binding-diagram}
\end{figure}
\subsection{Binding Diagram}
At zero pair momentum, we construct a phase diagram (Fig. \ref{fig:binding-diagram}) which shows where bound pairs form. States with nonzero angular momentum (i.e. $p$-, $d_{T_{2g}}$-, $d_{E_{g}}$- and $f$- states) are insensitive to $U$, as evident in Fig. \ref{fig:binding-diagram}.

In comparison to other lattices with lower coordination numbers, pairs require stronger attractions for their formation in the FCC lattice. This is due to the increase in kinetic energy  with coordination number. The critical $U$ or $V$ required for binding can be found analytically (refer to Appendix \ref{appendix:uv_derivation} for more details) via
\begin{equation}\label{eqn:critical_attraction}
    V_{c}^{s}(U)\leq \frac{UL_{0}-1}{UL_{0}\mathcal{C}-\mathcal{C}-12UL_{1}^2}
\end{equation}

\noindent where $L_{0}$=$-\sqrt{3}K_{0}^2/(8\pi^2t)$, $L_{1}$=$1/(24t)$+$L_{0}$, $\mathcal{C}$=$12L_0$+$1/(2t)$ and $K_{0}$=$K\left(\frac{\sqrt{3}-1}{2\sqrt{2}}\right)$=$1.598142\dots$ is the complete elliptic integral of the first kind. Note that Eqn. (\ref{eqn:critical_attraction}) only holds when $\Pvec=0$.

The required potential to create bound onsite pairs with no intersite interaction is $U_{c}^{\rm Hub.}$($V$=0) $\approx$ $-1.4874W$ (the negative Hubbard model, $V$=0). This is a slightly greater attraction relative to the simple cubic \cite{Davenport2012} and body-centred cubic \cite{adebanjo2021fermion} lattices. Equation
(\ref{eqn:critical_attraction}) also yields the critical attraction $V_{c}^{s}$($U$=0)$\approx$ $-0.4836W$. At infinite intersite repulsion, the onsite $s$-state is guaranteed to form if $U_{c}^{\rm Hub.}$($V$$\rightarrow$$+\infty$)$\leq$$-2W$. Also, $V_{c}^{s}$($+\infty$)=$-0.6502W$. The non-$s$ pairs have critical intersite binding strength $V_{c}^p$ $\approx$$-1.3586W$, $V_{c}^{d_{T_{2g}}}$$\approx$$-1.8945W$, $V_{c}^{d_{E_{g}}}$$\approx$$ -2.2342W$, $V_{c}^{f}$$\approx$$-2.2847W$.\\
\begin{table}[h!]
    {\renewcommand{\arraystretch}{1.5}
    \begin{tabular}{|l|c|c| c| c|}
    \hline
         \bf{Pairing} & \bf{Binding} & \multicolumn{3}{c|}{{\bf Minimum attraction required }}\\ \cline{3-5}
         \bf{symmetry} & \bf{parameter} & SC & BCC & FCC \\ \hline
         \multirow{4}{*}{$s$-wave} &  $U$($V$=0) & $-1.3189W_{\rm S}$ & $-1.4355W_{\rm B}$ & $-1.4874W_{\rm F}$  \\ \cline{2-5}
         & $U$($V$=+$\infty$) & $-2W_{\rm S}$ &  $-2W_{\rm B}$ &  $-2W_{\rm F}$ \\ \cline{2-5}
         & $V$($U$=0) & $-0.6455W_{\rm S}$ & $-0.6358W_{\rm B}$ & $-0.4836W_{\rm F}$ \\ \cline{2-5}
         & $V$($U$=+$\infty$) & $-0.9789W_{\rm S}$ & $-0.8858W_{\rm B}$ & $-0.6502W_{\rm F}$ \\ \hline
         $p$-wave ($A_{2u}$) & $V$ & $-1.5885W_{\rm S}$ & $-1.5828W_{\rm B}$ & $-1.3586W_{\rm F}$\\ \hline
         $d$-wave ($T_{2g}$) & $V$ &  & $-1.8804W_{\rm B}$ & $ -1.8945W_{\rm F}$\\ \hline
         $d$-wave ($E_{g}$) & $V$ & $-1.8034W_{\rm S}$ &  & $ -2.2342W_{\rm F}$\\ \hline
         $f$-wave ($T_{2u}$) & $V$ &  & $-1.9639W_{\rm B}$ & $ -2.2847W_{\rm F}$\\ \hline
    \end{tabular} }
    \caption{Comparing critical binding strengths at $\Pvec=0$ in 3D cubic lattices (simple cubic, body-centred cubic, face-centred cubic). $W_{\rm S}=6t$, $W_{\rm B}=8t$, and $W_{\rm F}=12t$ are the respective half-bandwidths. We note that there are more pairing states in the FCC lattice than the other lattices.}
    \label{table:compare_attractions_3D}
\end{table}

If  measured in terms of their respective bandwidths, there are similarities in the critical attractions needed to bind two fermions in the simple cubic, BCC and FCC lattices. The summary of this comparison is given in Table \ref{table:compare_attractions_3D}. 
\begin{figure*}[t!]
	\centering
	\includegraphics[width=\linewidth]{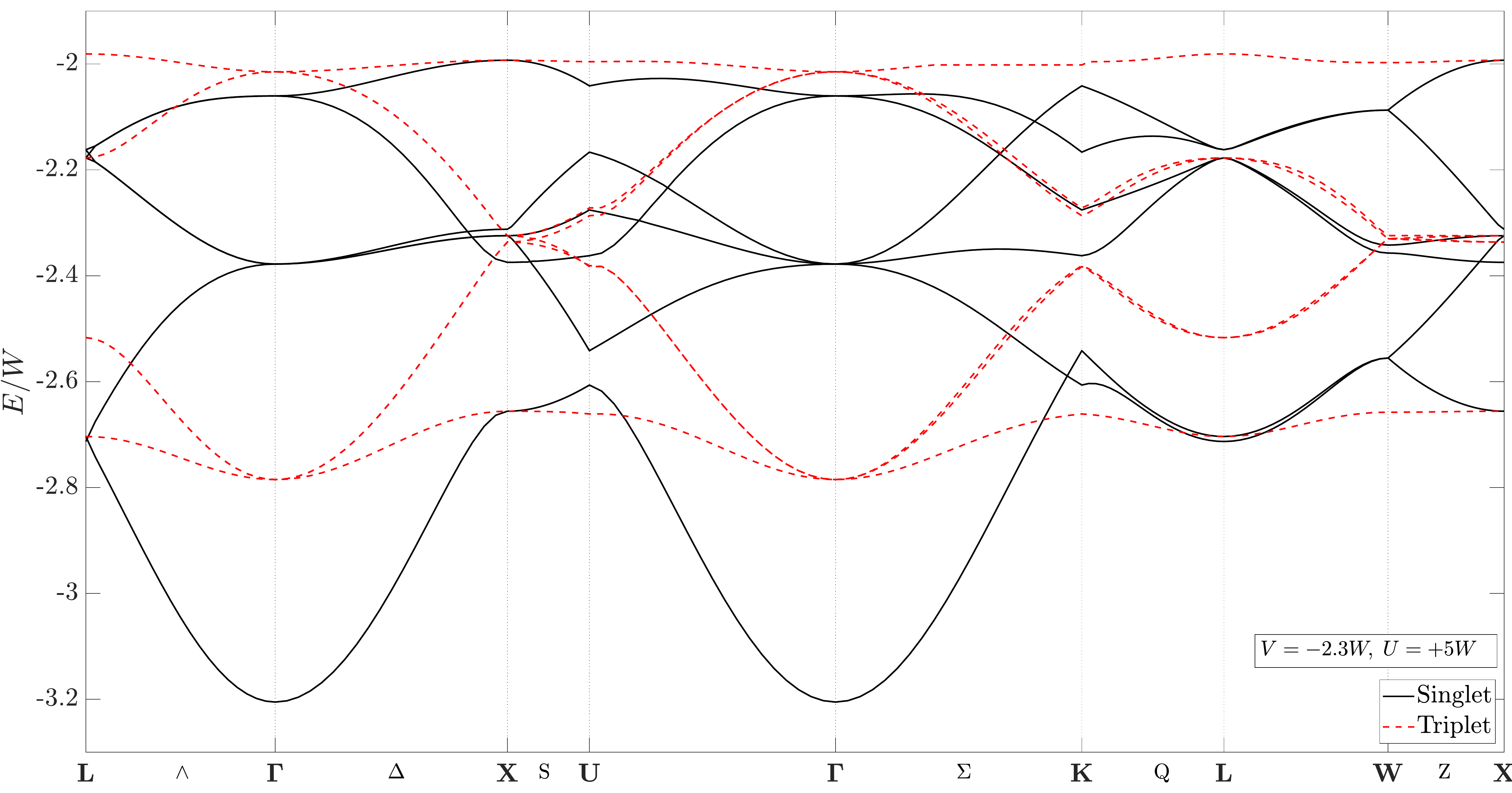}
	\caption{Dispersion curve of bound pairs at $U=+5W$ and $V=-2.3W$. Moving away from the $\Gamma$ point, degeneracies are lifted. Spin triplets are identified with a dashed (red) curve and spin singlets with solid (black) curve.}
	\label{fig:fcc_dispersion_plot}
\end{figure*}
\begin{figure*}
	\centering
	\includegraphics[width=85mm]{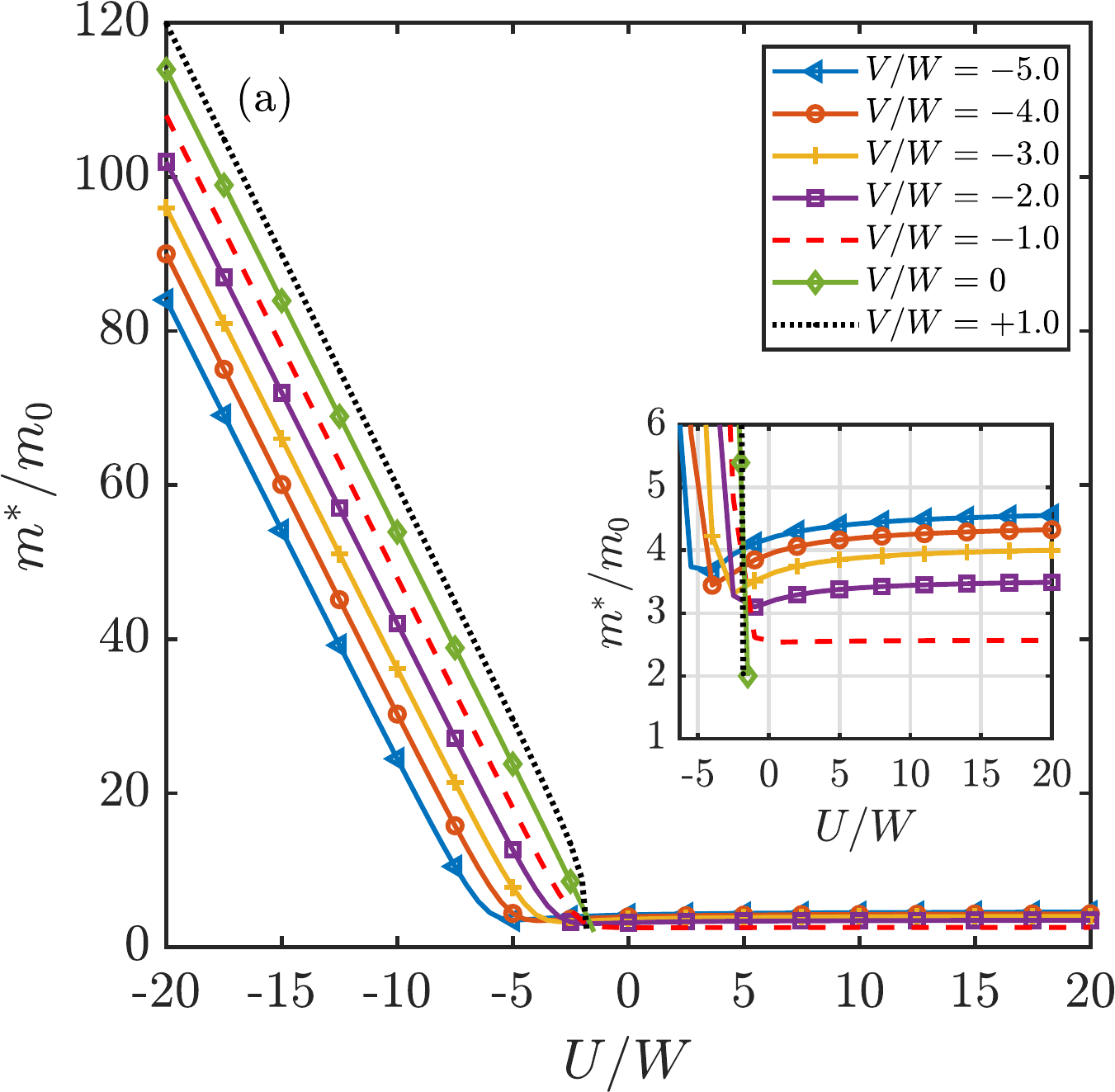}\vspace{.08em}
    \includegraphics[width=85mm]{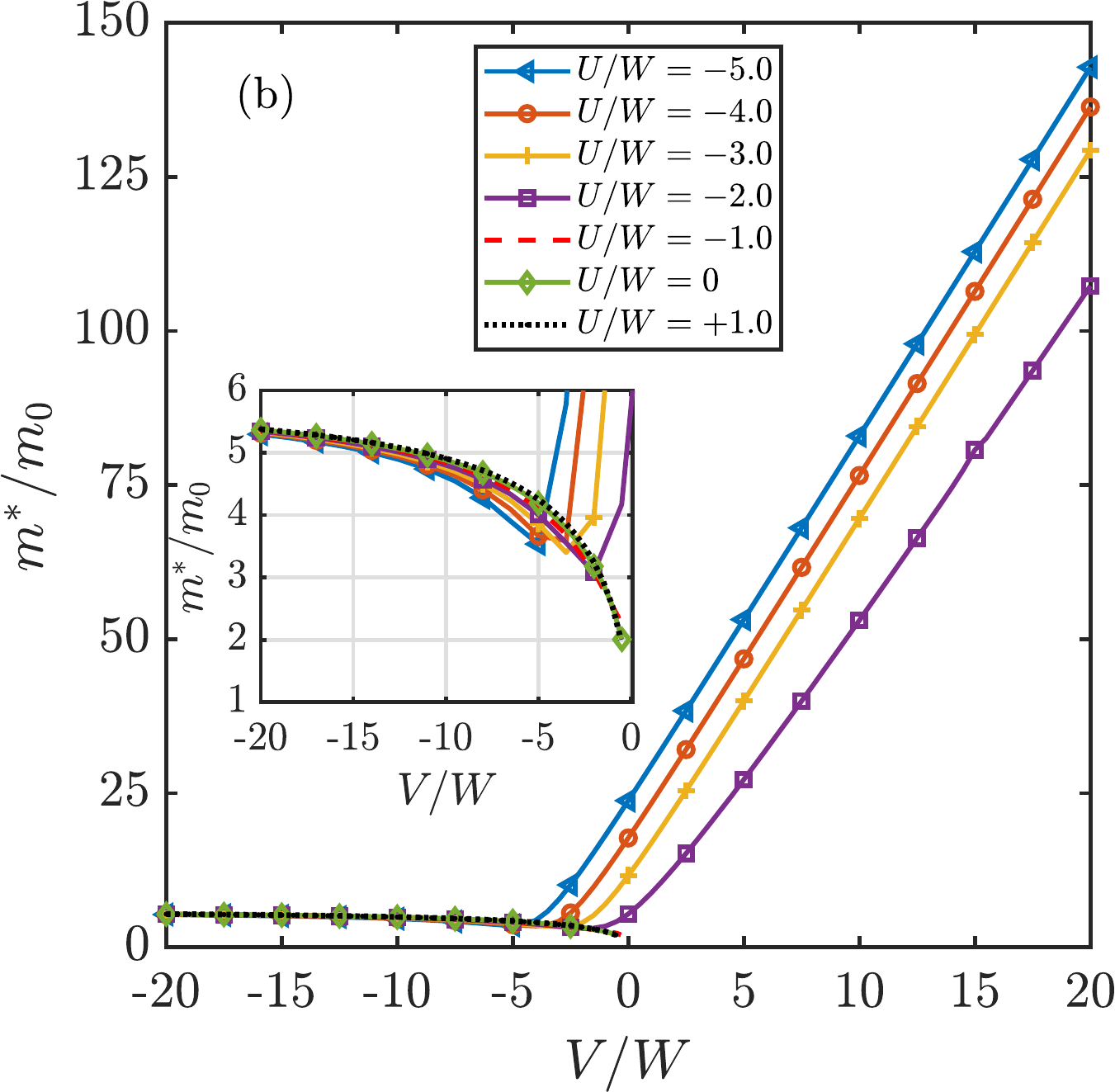}
	\caption{[Color online] Dependence of the pair mass on (a) $U$, (b) $V$. Pair mass increases with large and attractive $U$. In the limit $V\rightarrow-\infty$, the geometry of the FCC lattice means the pair mass tends to a value six times heavier than a single particle (inset of panel (b)).}
	\label{pair mass fcc}
\end{figure*}
\subsection{Dispersion}
The pair energy at non-zero momentum is needed to estimate the pair mass. The dispersion of singlet and triplet states across the FCC Brillouin zone (BZ) for $U$=$+5W$ (repulsive) and $V$=$-2.3W$ (attractive) is shown in Fig. \ref{fig:fcc_dispersion_plot}. To observe the dispersion of the excited states ($d_{E_{g}}$ and $f$), an intersite interaction stronger than the bandwidth ($-2W$) is required (evident from Table \ref{table:compare_attractions_3D}). Away from the $\Gamma$ point, degeneracies are lifted and there are mixing and crossing of states, making it difficult to specify pair symmetries. However, our method guarantees unambiguous classification of singlets and triplets. The overall form of the dispersions has been confirmed at strong coupling with perturbation theory calculations (see Appendix \ref{appendix:superlightmass}).

\begin{figure*}
	\centering
	\includegraphics[width=85mm]{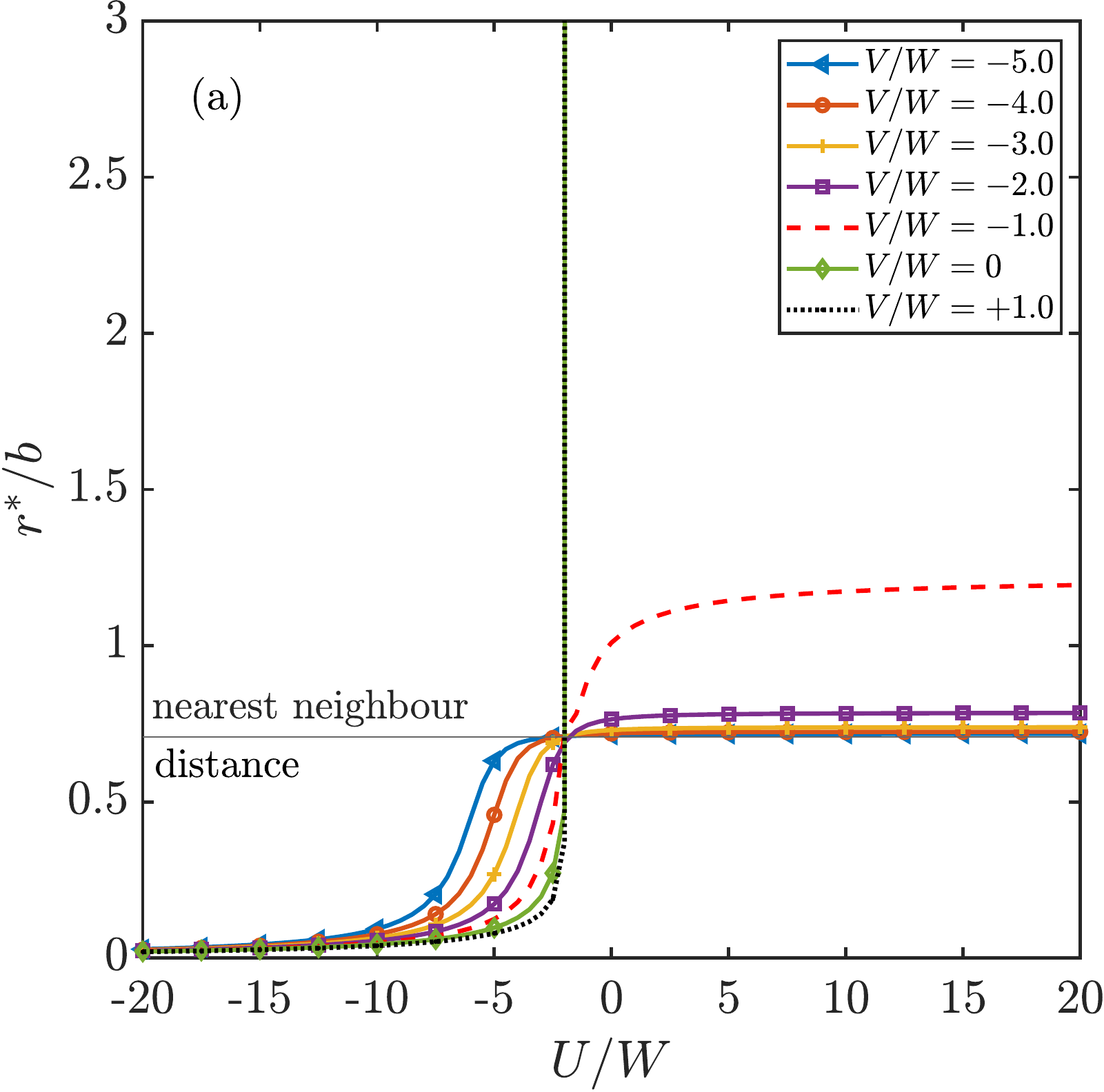}
	\includegraphics[width=85mm]{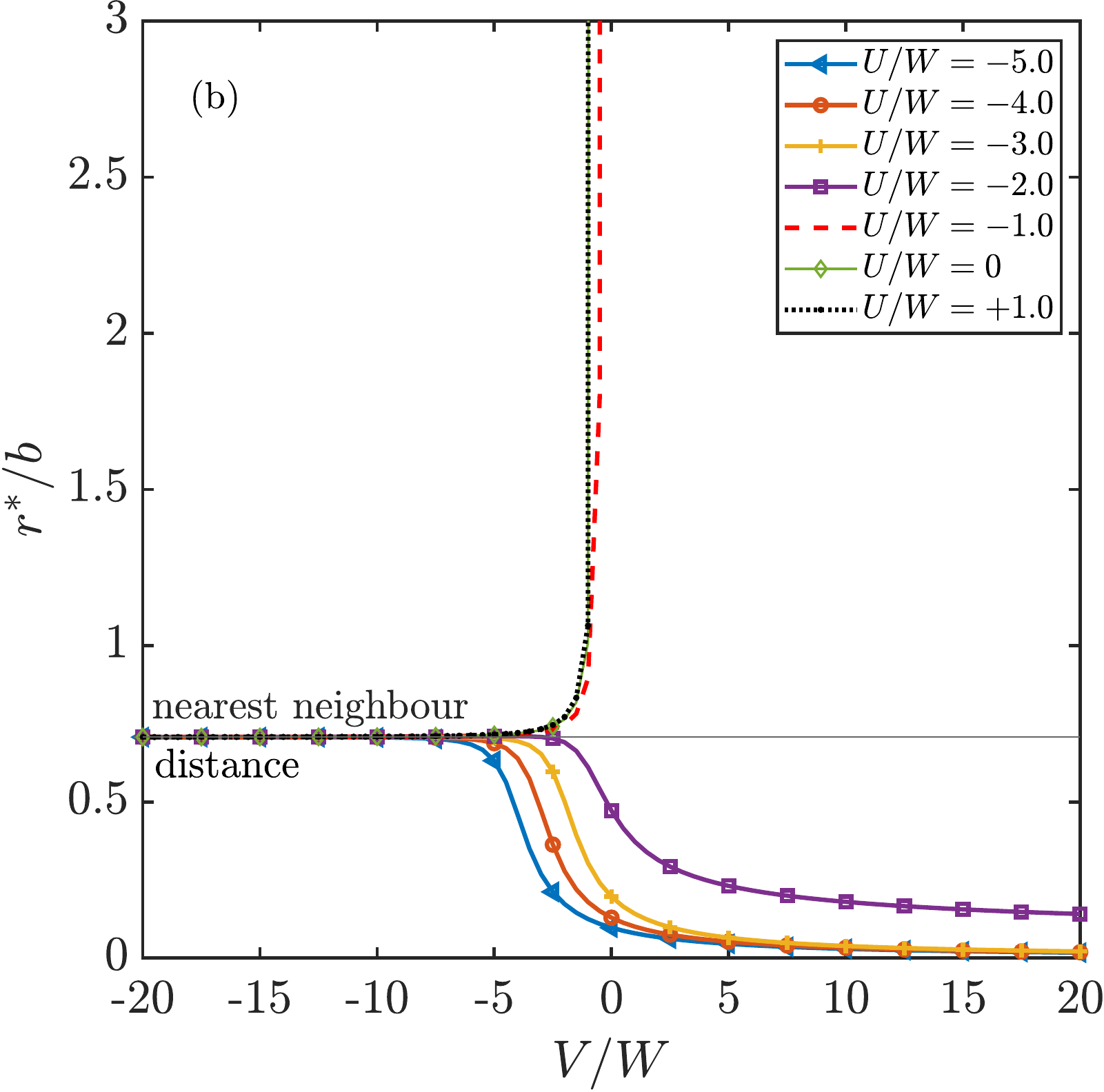}
	\caption{[Color online] Radius in units of the lattice constant for (a) onsite interaction at different $V$, (b) intersite interaction for various $U$. The radius diverges at the binding threshold. Intersite pairs form when $V$ is attractive and dominates, but when attractive Hubbard $U$ dominates, the formation of an onsite pair is favoured.}
	\label{fig:pair_size_fcc}
\end{figure*}
\begin{figure*}
	\centering
	\includegraphics[width=175mm]{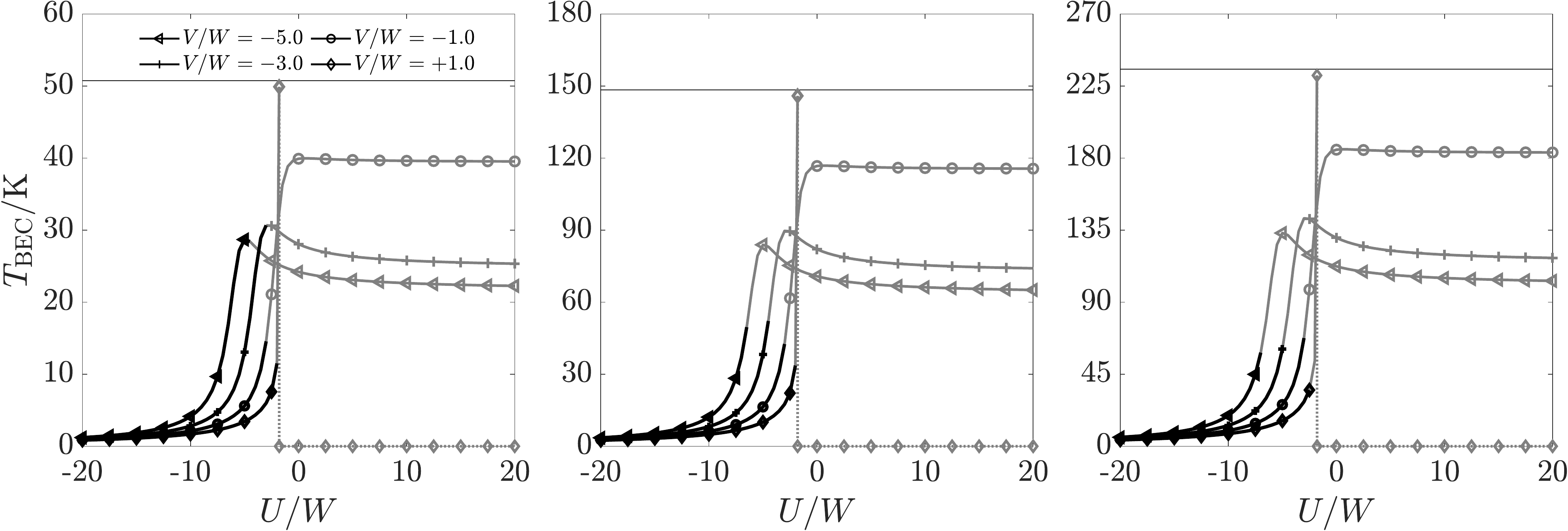}
	\includegraphics[width=175mm]{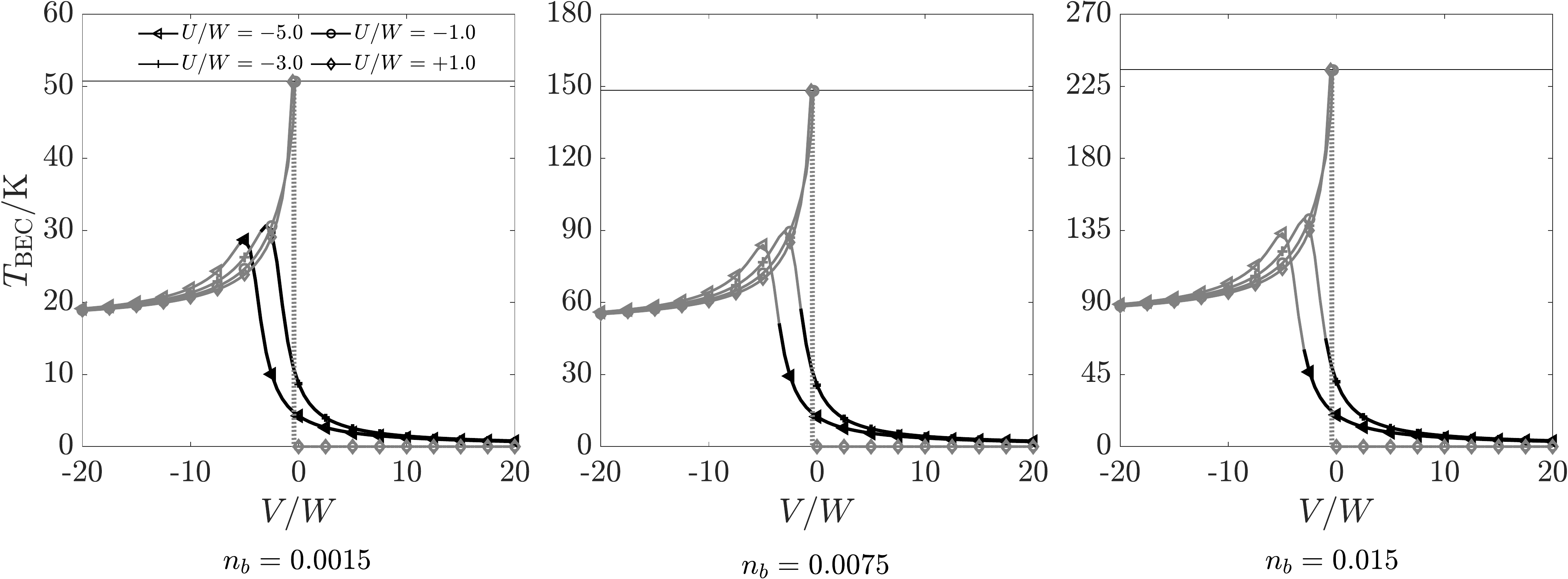}
	\caption{Plots of BEC transition temperature for bound pairs in the dilute limit. The number of pairs per site $n_b$ increases from left to right. Dark lines are used if $n_{b}$ satisfies the pair overlap condition. We observe a region of constant $T_{\rm BEC}$ in the superlight regimes. The horizontal lines show $T_{\rm BEC}$ for pairs of mass $m^* = 2m_{0}$. The dotted regions (abrupt decrease to zero transition temperature) indicate $T_\Delta < T_{\rm BEC}$ (so there are no preformed pairs).}
	\label{fig:tc_plots}
\end{figure*}
\subsection{Pair Mass}\label{sec:mass}
 The effective mass is calculated from the second derivative of the pair energy dispersion as
\begin{equation}
    [m^{*}_i]^{-1}=\frac{1}{\hbar^2}\frac{\partial^2 E}{\partial P_{i}^2}\;\;\;\;\;\;, 
\end{equation}

\noindent and masses can be seen in Fig. \ref{pair mass fcc}. 

In some lattices (e.g. rectangular ladder \cite{hague2007superlighta}, simple cubic \cite{Davenport2012}, and BCC \cite{adebanjo2021fermion}), a superlight itinerant state is formed only when $U$ and $V$ are nearly equal and are both attractive, which enhances the mobility of the pair via a single hop. However, for superlight pairs in staggered ladder  \cite{hague2007superlightb} and triangular lattices  \cite{hague2008_sing_trip_bip_triangular}, attractive onsite attraction is not required: a condition which is closer to physical systems which typically have onsite repulsion. The FCC lattice, due to its structure, belongs to the latter group.

In the limit where the intersite interaction is attractive and dominant $V$$\rightarrow$$-\infty$, the pair mass (Fig. \ref{pair mass fcc}b) tends to $m^*=6m_0$ (see Appendix \ref{appendix:superlightmass} for perturbation theory calculations in the large attractive $V$ limit) since the movement of the pair only depends linearly on $t$. On the other hand, in the limit where the onsite attraction is largely dominant $U$$\rightarrow$$-\infty$ and $V$ is small or repulsive, the mass of the bound pair increases with $|U|$. In this case, the pair movement is second-order in the hopping parameter $t$, with pair mass increasing with $|U|$, because it is necessary to hop through a higher energy intersite state for motion to occur.

\subsection{Pair Radius}

The effective radius is obtained from the expression
\begin{equation}\label{pair_radius_equation}
    \langle r^*\rangle=\sqrt{\frac{\sum\limits_{\nvec}\nvec^2\Psi^*(\nvec_{1},\nvec_{2})\Psi(\nvec_{1},\nvec_{2}) }{\sum\limits_{\nvec}\Psi^*(\nvec_{1},\nvec_{2})\Psi(\nvec_{1},\nvec_{2})}} \;\;\;,
\end{equation}
\noindent where $\nvec = |\nvec_{1}-\nvec_{2}|$. The radius is plotted in Fig. \ref{fig:pair_size_fcc}. 

The size of the pair diverges near the threshold energy ($E\rightarrow E^{\rm Th}$). This is because the pair is weakly bound and the pair wave function spreads over a distant lattice sites. If the particles unbind, the size becomes infinite. When $V$ is strongly attractive and dominates over $U$, we see the formation of a local intersite pair, with size on the order of the lattice parameter. In contrast, if the Hubbard attraction is very strong and dominant, an onsite pair is formed.

\subsection{Bose-Einstein Condensation}

We estimate BEC transition temperatures for pairs in an FCC lattice, obtained from the Bose integral as,
\begin{align}
	T_{\rm BEC} & = \frac{3.31\hbar^{2}}{m_{b}^{*}k_{B}} \left( \frac{n_{b}}{\Omega_{\rm site}} \right)^{2/3}
\end{align}
where $\Omega_{\rm site}= b^3/4$ is the volume of the Wigner--Seitz cell for the FCC lattice  and $n_{b}$ is the number of pairs per lattice site. We use lattice constant $b=14.24\, \si{\angstrom}$ as an example (consistent with the FCC fullerides, although we note that electrons are not dilute in fullerides and further manipulations would be required to derive a $UV$ model for such materials). Transition temperatures for fixed $n_{b}$ are plotted in Fig. \ref{fig:tc_plots}. For BEC to take place at $T_{\rm BEC}$, pairs must exist above $T_{\rm BEC}$, such that $T_{\rm BEC}<T_{\Delta}$, where $T_{\Delta}=\Delta/k_{B}$ is the pairing temperature and binding energy $\Delta=2\varepsilon_{0}-E_{0}$. The value $t=0.04$ eV (consistent with fulleride superconductors  \cite{gunnarsson2004alkali}) is used to set the energy scale. The maximum $n_{b}$ for which $T_{\rm BEC}$ is consistent with the effective mass approximation is estimated to be approximately 0.015. We require that pairs of radius $R'=\alpha r^{*}$ can fit into space without overlapping, so $n_{b}16 R^{\prime 3}/3<1$. Selecting $\alpha=5$ is suggested to minimise the overlap of wave functions of different pairs to $\sim 1\%$.
For $n_b=0.0015$, $T_{\rm BEC}\lesssim 30$ K and for $n_b=0.015$ $T_{\rm BEC}\lesssim 70$ K.

\section{Discussion and conclusions}\label{sec:discussion}
This paper explored the formation and properties of fermion pairs in an FCC lattice. The binding diagram, pair energy, effective mass and radius were calculated, and BEC transition temperatures were estimated. When the intersite attraction is large and $U$ is repulsive, leading to strongly bound intersite pairs, the bound pair was found to be six times heavier than a single particle mass, since geometric properties of FCC lattices allow motion of strongly bound intersite pairs as a first order effect. Infinitely heavy pairs are found on the FCC lattice for $U\rightarrow-\infty$.

Intersite pair motion on FCC lattices differs qualitatively from other 3D cubic lattices, where either next-nearest neighbour hopping or attractive Hubbard $U\sim V$ would be required to form light states when $V$ is strong and attractive. While effective attractive $U$ could be created if a very strong electron-phonon interaction overcome the Hubbard $U$, and this would lead to very heavy pairs due to polaron effects. The geometric properties of FCC lattices which do not require the electron-phonon interaction to overcome the Hubbard $U$, could lead to pairs, created by intersite electron-phonon interactions (and other mechanisms) that are light, mobile and undergo a superconducting transition at high temperatures.
 
It is possible to measure the mass of carriers in cuprate superconductors using optical and Hall measurements. This shows that in the quasi-2D high temperature superconductors, typical carrier masses are three electron masses, so light carriers are not necessarily unique to FCC lattices, and can be found if the intersite attraction is weak \cite{jozef2017,padilla2005}. We note that superlight states may not correspond to the highest transition temperatures for FCC lattices, and we estimate slightly higher $T_{\rm BEC}$ for weakly bound pairs. What is unusual here is that the pair masses are small, \emph{both} when the intersite attraction is weak \emph{and} when it is strong, which might allow superconductivity to be found at higher temperatures in a wider range of the parameter space of the extended Hubbard model on FCC lattices (perhaps making it easier to find materials with the right properties for superconductivity). We also note that it might be possible to probe the carrier mass in FCC fullerides using similar optical and Hall measurements.

Optical lattices may offer an alternative way to measure the properties of superlight pairs, since the mass enhancement of 6 in the strong attraction limit is a feature of the lattice geometry. FCC optical lattices can be formed using arrangements of 4 laser beams \cite{lang2017weyl,yuan2003arrangements}. The Hubbard $U$ in such systems can be controlled via the Feshbach resonance. Intersite $V$ is formed using dressed Rydberg states, and can be controlled using the principal quantum number for the Rydberg states and the detuning from the Rydberg state. In practice, FCC optical lattices have large lattice constants so we expect the BEC temperature will be low. However it may be possible to observe superlight pairs in optical lattice experiments in the normal state.

Further investigations using other theoretical techniques would be of interest. The challenge is to find techniques that can deal with the FCC structures without losing the detail of the lattice. For small numbers of particles, path integral QMC could be used \cite{hague2007superlighta}, but ED would not be appropriate in 3D as lattice sizes would be limited to less than 4 sites across due to the $4^N$ growth in Hilbert space. Approaching the thermodynamic limit is a challenge. DMFT is most suitable for 3D systems, but the coarse graining of the Brillouin zone washes out details of the lattice, such that results for all 3D lattices would be qualitatively identical.  We note that transition temperatures predicted here are similar to those in FCC fulleride materials. While the fullerides are not dilute, we suggest that future work could also include a determination of an effective $UV$ Hamiltonian for light doping away from half filling in such systems.

\bibliographystyle{unsrt}
\bibliography{References}

\appendix


%
%

%

\section{$UV$ model on face-centred cubic lattice}\label{appendix:uv_derivation}

\subsection{Schr\"odinger equation}\label{BCC_singlets_and_triplets}
The (anti-)symmetrised Schr\"odinger equation can be written as,
\begin{equation}\label{total_WF_appendix}
(E-\varepsilon_{\kvec_{1}}-\varepsilon_{\kvec_{2}})\phi_{\kvec_{1}\kvec_{2}}^{\pm}=\frac{1}{N}\sideset{}{'} \sum_{\qvec\avec_{\pm}}\hat{V}_{\avec_{\pm}}\;\Big\{e^{i(\qvec-\kvec_{1})\,\avec_{\pm}} \pm e^{i(\qvec - \kvec_{2})\,\avec_{\pm}} \Big\}\;\phi_{\qvec,\kvec_{1}+\kvec_{2}-\qvec}^{\pm}
\end{equation}
The prime in the summation means that a factor of $\frac{1}{2}$ must be included for the case $\avec_{+}=0$. Spin singlets have solutions belonging to the symmetrised Schr\"odinger equation while spin triplets can be found from the anti-symmetrised version of Eqn. (\ref{total_WF_appendix}).\\

\noindent We define the vectors for the singlets and triplets respectively as\\
$\{\avec_{+}\} = \{\avec^{+}_{0},\avec^{+}_{1},\avec^{+}_{2},\avec^{+}_{3},\avec^{+}_{4},\avec^{+}_{5},\avec^{+}_{6}\}= \{ 
(0,0,0), (\frac{1}{2},\frac{1}{2},0), (0,\frac{1}{2},\frac{1}{2}), (\frac{1}{2},0,\frac{1}{2}), (\frac{1}{2},-\frac{1}{2},0), (0,\frac{1}{2},-\frac{1}{2}),(-\frac{1}{2},0,\frac{1}{2})\}\quad$\\
$\{\avec_{-}\} = \{\avec^{-}_{1},\avec^{-}_{2},\avec^{-}_{3},\avec^{-}_{4},\avec^{-}_{5},\avec^{-}_{6}\}= \{(\frac{1}{2},\frac{1}{2},0), (0,\frac{1}{2},\frac{1}{2}), (\frac{1}{2},0,\frac{1}{2}), (\frac{1}{2},-\frac{1}{2},0), (0,\frac{1}{2},-\frac{1}{2}),(-\frac{1}{2},0,\frac{1}{2})\}$.\\
In this section, we set $b=1$.

\subsubsection{Symmetrised Sch\"odinger equation}
Using the vectors $\{\avec_{+}\}$ in Eqn. (\ref{total_WF_appendix}), we obtain
\begin{equation}\label{fccsingletsWF_appendix}
\begin{split}
& (E - \varepsilon_{\kvec_1} - \varepsilon_{\kvec_2})\phi_{\kvec_{1}\kvec_{2}}^{+} = \frac{1}{N}\sum_{\qvec}\bigg[\frac{1}{2}U(e^{i(\qvec-\kvec_{1})\avec^{+}_{0}}+e^{i(\qvec-\kvec_{2})\avec^{+}_{0}}) + V(e^{i(\qvec-\kvec_{1})\avec^{+}_{1}}+e^{i(\qvec-\kvec_{2})\avec^{+}_{1}})\\
& \:\:\:\:\;\:\:\:\:\; + V(e^{i(\qvec-\kvec_{1})\avec^{+}_{2}}+e^{i(\qvec-\kvec_{2})\avec^{+}_{2}}) + V(e^{i(\qvec-\kvec_{1})\avec^{+}_{3}}+e^{i(\qvec-\kvec_{2})\avec^{+}_{3}})+ V(e^{i(\qvec-\kvec_{1})\avec^{+}_{4}}+e^{i(\qvec-\kvec_{2})\avec^{+}_{4}}) \\
& \:\:\:\:\;\:\:\:\:\; + V(e^{i(\qvec-\kvec_{1})\avec^{+}_{5}}+e^{i(\qvec-\kvec_{2})\avec^{+}_{5}})+ V(e^{i(\qvec-\kvec_{1})\avec^{+}_{6}}+e^{i(\qvec-\kvec_{2})\avec^{+}_{6}})\bigg]\phi_{\qvec,\kvec_{1}+\kvec_{2}-\qvec}^{+}\\
& = \frac{1}{N}\sum_{\qvec}\bigg[U +V\;e^{i(\frac{q_{x}}{2}+\frac{q_{y}}{2})}(e^{-i \kvec_{1}\avec^{+}_{1}} + e^{-i \kvec_{2}\avec^{+}_{1}}) +V\; e^{i(\frac{q_{y}}{2}+\frac{q_{z}}{2})}(e^{-i \kvec_{1}\avec^{+}_{2}} + e^{-i \kvec_{2}\avec^{+}_{2}}) +V\; e^{i(\frac{q_{x}}{2}+\frac{q_{z}}{2})}(e^{-i \kvec_{1}\avec^{+}_{3}} + e^{-i \kvec_{2}\avec^{+}_{3}}) \\
& + V\; e^{i(\frac{q_{x}}{2}-\frac{q_{y}}{2})}(e^{-i \kvec_{1}\avec^{+}_{4}} + e^{-i \kvec_{2}\avec^{+}_{4}})+ V\; e^{i(\frac{q_{y}}{2}-\frac{q_{z}}{2})}(e^{-i \kvec_{1}\avec^{+}_{5}} + e^{-i \kvec_{2}\avec^{+}_{5}})+ V\; e^{i(-\frac{q_{x}}{2}+\frac{q_{z}}{2})}(e^{-i \kvec_{1}\avec^{+}_{6}} + e^{-i \kvec_{2}\avec^{+}_{6}}) \bigg]\phi_{\qvec,\kvec_{1}+\kvec_{2}-\qvec}^{+}
\end{split}
\end{equation}

The following basis functions can be used:
\begin{equation}\label{fccsingletfinalbasisfunction}
\begin{split}
& \Phi_{0}^{+}(\Pvec) = \frac{1}{N} \sum_{\qvec}\phi_{\qvec,\Pvec-\qvec}^{+} \;\mathrm{,}\;\;  \Phi_{1}^{+}(\Pvec) = \frac{1}{N} \sum_{\qvec}e^{i(\frac{q_x}{2}+\frac{q_y}{2})}\;\phi_{\qvec,\Pvec-\qvec}^{+} \;\mathrm{,}\;\;\Phi_{2}^{+}(\Pvec) = \frac{1}{N} \sum_{\qvec}e^{i(\frac{q_{y}}{2}+\frac{q_{z}}{2})}\;\phi_{\qvec,\Pvec-\qvec}^{+} \\
& \Phi_{3}^{+}(\Pvec) = \frac{1}{N} \sum_{\qvec}e^{i(\frac{q_{x}}{2}+\frac{q_{z}}{2})}\;\phi_{\qvec,\Pvec-\qvec}^{+} \;\;\;\mathrm{,}\;\;\;\;\Phi_{4}^{+}(\Pvec) = \frac{1}{N} \sum_{\qvec}e^{i(\frac{q_{x}}{2}-\frac{q_{y}}{2})}\;\phi_{\qvec,\Pvec-\qvec}^{+} \\
& \Phi_{5}^{+}(\Pvec) = \frac{1}{N} \sum_{\qvec}e^{i(\frac{q_{y}}{2}-\frac{q_{z}}{2})}\;\phi_{\qvec,\Pvec-\qvec}^{+} \;\;\;\mathrm{,}\;\;\;\;\Phi_{6}^{+}(\Pvec) = \frac{1}{N} \sum_{\qvec}e^{i(-\frac{q_{x}}{2}+\frac{q_{z}}{2})}\;\phi_{\qvec,\Pvec-\qvec}^{+}
\end{split}
\end{equation}
where $\Pvec=\kvec_1+\kvec_2$.
The wave function in Eqn. (\ref{fccsingletsWF_appendix}) expressed in terms of the basis functions Eqn. (\ref{fccsingletfinalbasisfunction}) is
\begin{equation}\label{singlet_all_phi_plus}
\begin{split}
& \phi_{\kvec_{1}\kvec_{2}}^{+} = \frac{1}{(E - \varepsilon_{\kvec_1} - \varepsilon_{\kvec_2})}\bigg\{ U\Phi_{0}^{+}(\Pvec) +V\;\Phi_{1}^{+}(\Pvec)(e^{-i\kvec_{1}\avec^{+}_{1}}+e^{-i\kvec_{2}\avec^{+}_{1}})+V\;\Phi_{2}^{+}(\Pvec)(e^{-i\kvec_{1}\avec^{+}_{2}}+e^{-i\kvec_{2}\avec^{+}_{2}}) \\
& \;\;\;\;\;\;\;\;\;\;\;\;\;\;+V\;\Phi_{3}^{+}(\Pvec)(e^{-i\kvec_{1}\avec^{+}_{3}}+e^{-i\kvec_{2}\avec^{+}_{3}}) +V\;\Phi_{4}^{+}(\Pvec)(e^{-i\kvec_{1}\avec^{+}_{4}}+e^{-i\kvec_{2}\avec^{+}_{4}})+V\;\Phi_{5}^{+}(\Pvec)(e^{-i\kvec_{1}\avec^{+}_{5}}+e^{-i\kvec_{2}\avec^{+}_{5}}) \\
& \;\;\;\;\;\;\;\;\;\;\;\;\;\; +V\;\Phi_{6}^{+}(\Pvec)(e^{-i\kvec_{1}\avec^{+}_{6}}+e^{-i\kvec_{2}\avec^{+}_{6}})\bigg\}
\end{split}
\end{equation}

Substituting Eqn. (\ref{singlet_all_phi_plus}) into each entry of Eqn. (\ref{fccsingletfinalbasisfunction}) and redefining $q_{j} = q_{j}^{'} + \frac{P_{j}}{2}$ leads to seven self-consistent equations. The first is
\begin{equation}
    \begin{split}
        & \tilde{\Phi}_{0}^{+}(\Pvec) = UL_{000}(\Pvec)\tilde{\Phi}_{0}^{+}(\Pvec)  + V\Big[L_{110}(\Pvec)+L_{\bar{1}\bar{1}0}(\Pvec)\Big]\tilde{\Phi}_{1}^{+}(\Pvec) + V\Big[L_{011}(\Pvec)+L_{0\bar{1}\bar{1}}(\Pvec)\Big]\tilde{\Phi}_{2}^{+}(\Pvec) \\
        & \;\;\;\;\;\;\;\;\;\;\;\;\;\; + V\Big[L_{101}(\Pvec)+L_{\bar{1}0\bar{1}}(\Pvec)\Big]\tilde{\Phi}_{3}^{+}(\Pvec) + V\Big[L_{1\bar{1}0}(\Pvec)+L_{\bar{1}10}(\Pvec)\Big]\tilde{\Phi}_{4}^{+}(\Pvec) \\
        & \;\;\;\;\;\;\;\;\;\;\;\;\;\;  + V\Big[L_{01\bar{1}}(\Pvec)+L_{0\bar{1}1}(\Pvec)\Big]\tilde{\Phi}_{5}^{+}(\Pvec)  +  V\Big[L_{\bar{1}01}(\Pvec)+L_{10\bar{1}}(\Pvec)\Big]\tilde{\Phi}_{6}^{+}(\Pvec)
    \end{split}
    \label{eqn:phitildezero}
\end{equation}
where $\tilde{\Phi}_{i}^{+}(\Pvec)=e^{\frac{-i}{2}(\Pvec\avec_{i}^{+})}\Phi_{i}^{+}$ (where $i=0,1,...6$), have phase factors which provides information about the centre-of-mass motion of the bound state. Furthermore, the $L$'s represent the Green's functions of the FCC lattice which are defined as
\begin{equation}\label{greens_function_appendix}
    L_{lmn}(\Pvec)=\frac{1}{N}\sum_{\qvec^{'}}\frac{e^{i(l\frac{q_{x}^{'}}{2}+m\frac{q_{y}^{'}}{2}+n\frac{q_{z}^{'}}{2})}}{E-\varepsilon_{\frac{\Pvec}{2}+\qvec^{'}}-\varepsilon_{\frac{\Pvec}{2}-\qvec^{'}}}=-\int_{-2\pi}^{2\pi}\int_{-2\pi}^{2\pi}\int_{-2\pi}^{2\pi}\frac{dq_{x}^{'} dq_{y}^{'} dq_{z}^{'}}{(4\pi)^{3}}\frac{\cos (l\frac{q_{x}^{'}}{2}+m\frac{q_{y}^{'}}{2}+n\frac{q_{z}^{'}}{2})}{|E|+\varepsilon_{\frac{\Pvec}{2}+\qvec^{'}}+\varepsilon_{\frac{\Pvec}{2}-\qvec^{'}}}
\end{equation}
Instead of writing a negative subscript along a coordinate, we place a bar above it to keep notation compact. For the present problem where interactions are only limited to nearest-neighbour distances, $l$, $m$ and $n$ take the values 0, $\pm$1, and $\pm$2. 

Combining all seven independent self-consistent equations for all spin-singlets at arbitrary momentum gives
\begin{equation}\label{appendix_singlet_sel_cons_eqn}
    \colvec[0.97]{
    UL_{000} & V[L_{110}+L_{\bar{1}\bar{1}0}] & V[L_{011}+L_{0\bar{1}\bar{1}}] & V[L_{101}+L_{\bar{1}0\bar{1}} ]  & V[L_{1\bar{1}0}+L_{\bar{1}10}] & V (L_{01\bar{1}}+L_{0\bar{1}1}] & V[L_{\bar{1}01}+L_{10\bar{1}}] \\
		UL_{110} & V[L_{000}+L_{220}] & V[L_{10\bar{1}} + L_{121}] & V[L_{01\bar{1}} + L_{211}] & V[L_{020} + L_{200}] & V[L_{101} + L_{12\bar{1}}] & V[L_{21\bar{1}} + L_{011}] \\
		UL_{011} & V[L_{\bar{1}01} + L_{121}] & V[L_{000} + L_{022}] & V[L_{\bar{1}10}+L_{112}] & V[L_{\bar{1}21} + L_{101}] & V[L_{002}+L_{020}] & V[L_{110}+L_{\bar{1}12}] \\
		UL_{101} & V[L_{0\bar{1}1} + L_{211}] & V[L_{1\bar{1}0} + L_{112}] & V[L_{000}+L_{202}] & V[L_{011} + L_{2\bar{1}1}] & V[L_{1\bar{1}2}+L_{110}] & V[L_{200} + L_{002}] \\
		UL_{1\bar{1}0} & V[L_{0\bar{2}0} + L_{200}] & V[L_{1\bar{2}\bar{1}} + L_{101}] & V[L_{0\bar{1}\bar{1}}+L_{2\bar{1}1}] & V[L_{000} + L_{2\bar{2}0} & V[L_{1\bar{2}1} + L_{10\bar{1}}] & V[L_{2\bar{1}\bar{1}} + L_{0\bar{1}1}] \\
		UL_{01\bar{1}} & V[L_{\bar{1}0\bar{1}} + L_{12\bar{1}}] & V[L_{00\bar{2}} + L_{0\bar{2}0}] & V[L_{\bar{1}1\bar{2}}+L_{110}] & V[L_{\bar{1}2\bar{1}}+L_{10\bar{1}}] & V[L_{000}+L_{02\bar{2}}] & V[L_{11\bar{2}}+L_{\bar{1}10}] \\
		UL_{\bar{1}01} & V[L_{\bar{2}\bar{1}1}+L_{011}] & V[L_{\bar{1}\bar{1}0}+L_{\bar{1}12}] & V[L_{\bar{2}00}+L_{002}] & V[L_{\bar{2}11}+L_{0\bar{1}1}] & V[L_{\bar{1}\bar{1}2}+L_{\bar{1}10}] & V[L_{000}+L_{\bar{2}02}]
    }\colvec[.85]{
    \tilde{\Phi}_{0}^{+}\\ \tilde{\Phi}_{1}^{+}\\ \tilde{\Phi}_{2}^{+}\\ \tilde{\Phi}_{3}^{+}\\ \tilde{\Phi}_{4}^{+} \\ \tilde{\Phi}_{5}^{+} \\ \tilde{\Phi}_{6}^{+}
    }
    =
    \colvec[.85]{
    \tilde{\Phi}_{0}^{+}\\ \tilde{\Phi}_{1}^{+}\\ \tilde{\Phi}_{2}^{+}\\ \tilde{\Phi}_{3}^{+}\\ \tilde{\Phi}_{4}^{+} \\ \tilde{\Phi}_{5}^{+} \\ \tilde{\Phi}_{6}^{+}}
\end{equation}
%
\subsubsection{Anti-symmetrised Schr\"odinder equation}
The anti-symmetrised equation is found by substituting $\{\mathbf{a_{-}}\}$ in Eqn. (\ref{total_WF_appendix}),
\begin{equation}\label{fcctripletsymmetrizedeqution}
\begin{split}
& (E - \varepsilon_{\kvec_1} - \varepsilon_{\kvec_2})\phi_{\kvec_{1}\kvec_{2}}^{-} = \frac{1}{N}\sum_{\qvec}\bigg[ V\;e^{i(\frac{q_{x}}{2}+\frac{q_{y}}{2})}(e^{-i \kvec_{1}\avec^{-}_{1}} - e^{-i \kvec_{2}\avec^{-}_{1}}) +V\; e^{i(\frac{q_{y}}{2}+\frac{q_{z}}{2})}(e^{-i \kvec_{1}\avec^{-}_{2}} - e^{-i \kvec_{2}\avec^{-}_{2}}) \\
& \;\;\;\;\;\;\;\;\;\;\;\;\; +  V\; e^{i(\frac{q_{x}}{2}+\frac{q_{z}}{2})}(e^{-i \kvec_{1}\avec^{-}_{3}} - e^{-i \kvec_{2}\avec^{-}_{3}}) + V\; e^{i(\frac{q_{x}}{2}-\frac{q_{y}}{2})}(e^{-i \kvec_{1}\avec^{-}_{4}} - e^{-i \kvec_{2}\avec^{-}_{4}}) \\
&  \;\;\;\;\;\;\;\;\;\;\;\;\; + V\; e^{i(\frac{q_{y}}{2}-\frac{q_{z}}{2})}(e^{-i \kvec_{1}\avec^{-}_{5}} - e^{-i \kvec_{2}\avec^{-}_{5}})+ V\; e^{i(-\frac{q_{x}}{2}+\frac{q_{z}}{2})}(e^{-i \kvec_{1}\avec^{-}_{6}} - e^{-i \kvec_{2}\avec^{-}_{6}}) \bigg]\phi_{\qvec,\kvec_{1}+\kvec_{2}-\qvec}^{-}
\end{split}
\end{equation}
Similar basis functions can be used to the singlet case:
\begin{equation}\label{appendix_triplet_phi}
\begin{split}
& \Phi_{1}^{-}(\Pvec) = \frac{1}{N} \sum_{\qvec}e^{i(\frac{q_x}{2}+\frac{q_y}{2})}\;\phi_{\qvec,\Pvec-\qvec}^{-} \;\mathrm{,}\;\;\Phi_{2}^{-}(\Pvec) = \frac{1}{N} \sum_{\qvec}e^{i(\frac{q_{y}}{2}+\frac{q_{z}}{2})}\;\phi_{\qvec,\Pvec-\qvec}^{-} \\
& \Phi_{3}^{-}(\Pvec) = \frac{1}{N} \sum_{\qvec}e^{i(\frac{q_{x}}{2}+\frac{q_{z}}{2})}\;\phi_{\qvec,\Pvec-\qvec}^{-} \;\;\;\mathrm{,}\;\;\;\;\Phi_{4}^{-}(\Pvec) = \frac{1}{N} \sum_{\qvec}e^{i(\frac{q_{x}}{2}-\frac{q_{y}}{2})}\;\phi_{\qvec,\Pvec-\qvec}^{-} \\
& \Phi_{5}^{-}(\Pvec) = \frac{1}{N} \sum_{\qvec}e^{i(\frac{q_{y}}{2}-\frac{q_{z}}{2})}\;\phi_{\qvec,\Pvec-\qvec}^{-} \;\;\;\mathrm{,}\;\;\;\;\Phi_{6}^{-}(\Pvec) = \frac{1}{N} \sum_{\qvec}e^{i(-\frac{q_{x}}{2}+\frac{q_{z}}{2})}\;\phi_{\qvec,\Pvec-\qvec}^{-}
\end{split}
\end{equation}
leading to
\begin{equation}
    \begin{split}
        & \tilde{\Phi}_{1}^{-}(\Pvec) =  V\Big[L_{000}-L_{220}\Big]\tilde{\Phi}_{1}^{-}(\Pvec) + V\Big[L_{10\bar{1}}-L_{121}\Big]\tilde{\Phi}_{2}^{-}(\Pvec)  + V\Big[L_{01\bar{1}}-L_{211}\Big]\tilde{\Phi}_{3}^{-}(\Pvec) + V\Big[L_{020}-L_{200}\Big]\tilde{\Phi}_{4}^{-}(\Pvec)  \\
        & \;\;\;\;\;\;\;\;\;\;\;\;\;\;\;\; + V\Big[L_{101}-L_{12\bar{1}}\Big]\tilde{\Phi}_{5}^{-}(\Pvec) + V\Big[L_{21\bar{1}}-L_{011}\Big]\tilde{\Phi}_{6}^{-}(\Pvec)
    \end{split}
    \label{eqn:phitildeminusone}
\end{equation}
The combined self-consistent equations for all triplets are 
\begin{equation}\label{appendix_triplet_sel_cons_eqn}
    \colvec[.99]{
    	V[L_{000}-L_{220}] & V[L_{10\bar{1}} - L_{121}] & V[L_{01\bar{1}} - L_{211}] & V[L_{020} - L_{200}] & V[L_{101} - L_{12\bar{1}}] & V[L_{21\bar{1}} - L_{011}] \\
	V[L_{\bar{1}01} - L_{121}] & V[L_{000} - L_{022}] & V[L_{\bar{1}10} - L_{112}] & V[L_{\bar{1}21} - L_{101}] & V[L_{002} - L_{020}] & V[L_{110}-L_{\bar{1}12}] \\
	V[L_{0\bar{1}1} -L_{211}] & V[L_{1\bar{1}0} - L_{112}] & V[L_{000}-L_{202}] & V[L_{011} - L_{2\bar{1}1}] & V[L_{1\bar{1}2}-L_{110}] & V[L_{200} - L_{002}] \\
	V[L_{0\bar{2}0} - L_{200}] & V[L_{1\bar{2}\bar{1}} - L_{101}] & V[L_{0\bar{1}\bar{1}}-L_{2\bar{1}1}] & V[L_{000} - L_{2\bar{2}0} & V[L_{1\bar{2}1} - L_{10\bar{1}}] & V[L_{2\bar{1}\bar{1}} - L_{0\bar{1}1} ] \\
	V[L_{\bar{1}0\bar{1}} - L_{12\bar{1}}] & V[L_{00\bar{2}} - L_{0\bar{2}0}] & V[L_{\bar{1}1\bar{2}}-L_{110}] & V[L_{\bar{1}2\bar{1}} - L_{10\bar{1}}] & V[L_{000}-L_{02\bar{2}}] & V[L_{11\bar{2}}-L_{\bar{1}10}] \\
	V[L_{\bar{2}\bar{1}1}-L_{011}] & V[L_{\bar{1}\bar{1}0}-L_{\bar{1}12}] & V[L_{\bar{2}00}-L_{002}] & V[L_{\bar{2}11}-L_{0\bar{1}1}] & V[L_{\bar{1}\bar{1}2}-L_{\bar{1}10}] & V[L_{000}-L_{\bar{2}02}]
	}
	\colvec[.85]{
	\tilde{\Phi}_{1}^{-}\\ \tilde{\Phi}_{2}^{-}\\ \tilde{\Phi}_{3}^{-}\\ \tilde{\Phi}_{4}^{-} \\ \tilde{\Phi}_{5}^{-} \\ \tilde{\Phi}_{6}^{-}
	}
	=
	\colvec[.85]{
	\tilde{\Phi}_{1}^{-}\\ \tilde{\Phi}_{2}^{-}\\ \tilde{\Phi}_{3}^{-}\\ \tilde{\Phi}_{4}^{-} \\ \tilde{\Phi}_{5}^{-} \\ \tilde{\Phi}_{6}^{-}}
\end{equation}
%
%
\subsection{Pair energy for $\Gamma$ point}
At the $\Gamma$ point,
\begin{equation}
\begin{split}
& L_{lmn}(0)=\frac{1}{N}\sum_{\qvec^{'}}\frac{e^{i(l\frac{q_{x}^{'}}{2}+m\frac{q_{y}^{'}}{2}+n\frac{q_{z}^{'}}{2})}}{E-2\varepsilon_{\qvec^{'}}} \\
& =-\int_{-2\pi}^{2\pi}\int_{-2\pi}^{2\pi}\int_{-2\pi}^{2\pi}\frac{dq_{x}^{'} dq_{y}^{'} dq_{z}^{'}}{(4\pi)^{3}}\frac{\cos (l\frac{q_{x}^{'}}{2})\;\cos(m\frac{q_{y}^{'}}{2})\;\cos(n\frac{q_{z}^{'}}{2})}{|E|- 8t\Big\{\cos(\frac{q_{x}^{'}}{2})\cos(\frac{q_{y}^{'}}{2}) + \cos(\frac{q_{x}^{'}}{2})\cos(\frac{q_{z}^{'}}{2}) + \cos(\frac{q_{y}^{'}}{2})\cos(\frac{q_{z}^{'}}{2})\Big\}} \\ & =-\frac{1}{(2\pi)^3} \int_{-\pi}^{\pi}\int_{-\pi}^{\pi}\int_{-\pi}^{\pi}\frac{\cos(lq_{x}^{''})\cos(mq_{y}^{''})\cos(nq_{z}^{''})}{|E|-8t\Big\{\cos(q_{x}^{''})\cos(q_{y}^{''}) + \cos(q_{x}^{''})\cos(q_{z}^{''}) + \cos(q_{y}^{''})\cos(q_{z}^{''})\Big\}}dq_{x}^{''}\,dq_{y}^{''}dq_{z}^{''}\;\;: \;\;\;(q_{j}^{''} = \frac{q_{j}^{'}}{2})
\end{split}
\end{equation}
Because some of the Green's functions are identical due to symmetry properties \cite{morita1975use}, we may then use the simplifications below

\begin{equation}\label{all_gf_at_gamma_point}
\begin{split}
& L_{000} = L_{0} \\
& L_{110} =L_{101} = L_{011}=L_{\bar{1}01}=L_{0\bar{1}1}=L_{10\bar{1}}=L_{\bar{1}0\bar{1}}=L_{0\bar{1}\bar{1}}=L_{01\bar{1}}=L_{\bar{1}10}=L_{1\bar{1}0}=L_{\bar{1}\bar{1}0}\equiv L_{1} \\
& L_{220} = L_{022} = L_{202}=L_{2\bar{2}0}=L_{02\bar{2}}=L_{\bar{2}02}\equiv L_{2} \\
& L_{200} = L_{020} =L_{002} =L_{\bar{2}00}=L_{0\bar{2}0}=L_{00\bar{2}}\equiv L_{3} \\
& L_{211} = L_{121} = L_{112}=L_{\bar{2}11}=L_{1\bar{2}1}=L_{11\bar{2}} =L_{\bar{1}\bar{1}2} =...\equiv L_{4}
\end{split}
\end{equation}
to modify our dispersion matrices (\ref{appendix_singlet_sel_cons_eqn}) and (\ref{appendix_triplet_sel_cons_eqn}) to obtain (at the $\Gamma$ point, $\tilde{\Phi}_{i}^{\pm}=\Phi_{i}^{\pm}$)
:
\begin{gather}
\underbrace{\begin{pmatrix}
	UL_{0} & 2VL_{1} & 2VL_{1} & 2VL_{1} & 2VL_{1} & 2VL_{1} & 2VL_{1} \\
	UL_{1} & V[L_{0} + L_{2}] & V[L_{1} + L_{4}] & V[L_{1} + L_{4}] & 2VL_{3} & V[L_{1} + L_{4}] & V[L_{4} + L_{1}] \\
	UL_{1} & V[L_{1} + L_{4}] & V[L_{0} + L_{2}] & V[L_{1}+L_{4}] & V[L_{4} + L_{1}] & 2VL_{3} & V[L_{1} + L_{4}]\\
	UL_{1} & V[L_{1} + L_{4}] & V[L_{1} + L_{4}] & V[L_{0}+L_{2}] & V[L_{1} + L_{4}] & V[L_{4} + L_{1}] & 2VL_{3} \\
	UL_{1} & 2VL_{3} & V[L_{4} + L_{1}] & V[L_{1}+L_{4}] & V[L_{0} + L_{2}] & V[L_{4} + L_{1}] & V[L_{4} + L_{1}] \\
	UL_{1} & V[L_{1}+L_{4}] & 2VL_{3} & V[L_{4}+L_{1}] & V[L_{4} + L_{1}] & V[L_{0} + L_{2}] & V[L_{4} + L_{1}] \\
	UL_{1} & V[L_{4}+L_{1}] & V[L_{1}+L_{4}] & 2VL_{3} & V[L_{4} + L_{1}] & V[L_{4} + L_{1}] & V[L_{0} + L_{2}]
	\end{pmatrix}}_{\hat{L}_{\rm singlet}}
\underbrace{\begin{pmatrix}
	\Phi_{0}^{+}\\ \Phi_{1}^{+}\\ \Phi_{2}^{+}\\ \Phi_{3}^{+}\\ \Phi_{4}^{+} \\ \Phi_{5}^{+}\\ \Phi_{6}^{+}
	\end{pmatrix}}_{\hat{\Phi}_{\rm singlet}}
=
\underbrace{\begin{pmatrix}
	\Phi_{0}^{+}\\ \Phi_{1}^{+}\\ \Phi_{2}^{+}\\ \Phi_{3}^{+}\\ \Phi_{4}^{+} \\ \Phi_{5}^{+}\\ \Phi_{6}^{+}
	\end{pmatrix}}_{\hat{\Phi}_{\rm singlet}}
\end{gather}
\begin{gather}
\underbrace{\begin{pmatrix}
	V[L_{0} - L_{2}] & V[L_{1} - L_{4}] & V[L_{1} - L_{4}] & 0 & V[L_{1} - L_{4}] & V[L_{4} - L_{1}] \\
	V[L_{1} - L_{4}] & V[L_{0} - L_{2}] & V[L_{1} - L_{4}] & V[L_{4} - L_{1}] & 0 & V[L_{1} - L_{4}]\\
	V[L_{1} - L_{4}] & V[L_{1} - L_{4}] & V[L_{0} - L_{2}] & V[L_{1} - L_{4}] & V[L_{4} - L_{1}] & 0 \\
	0 & V[L_{4} - L_{1}] & V[L_{1} - L_{4}] & V[L_{0} - L_{2}] & V[L_{4} - L_{1}] & V[L_{4} - L_{1}] \\
	V[L_{1} - L_{4}] & 0 & V[L_{4} - L_{1}] & V[L_{4} - L_{1}] & V[L_{0} - L_{2}] & V[L_{4} - L_{1}] \\
	V[L_{4} - L_{1}] & V[L_{1} - L_{4}] & 0 & V[L_{4} - L_{1}] & V[L_{4} - L_{1}] & V[L_{0} - L_{2}]
	\end{pmatrix}}_{\hat{L}_{\rm triplet}}
\underbrace{\begin{pmatrix}
	\Phi_{1}^{-}\\ \Phi_{2}^{-}\\ \Phi_{3}^{-}\\ \Phi_{4}^{-} \\ \Phi_{5}^{-} \\ \Phi_{6}^{-}
	\end{pmatrix}}_{\hat{\Phi}_{\rm triplet}}
=
\underbrace{\begin{pmatrix}
	\Phi_{1}^{-}\\ \Phi_{2}^{-}\\ \Phi_{3}^{-}\\ \Phi_{4}^{-} \\ \Phi_{5}^{-} \\ \Phi_{6}^{-}
	\end{pmatrix}}_{\hat{\Phi}_{\rm triplet}}
\end{gather}
%
We may rewrite the matrices above as
\begin{equation}
	\hat{L}_{s,t}\,\hat{\Phi}_{s,t}= \lambda_{s,t}\,\hat{\Phi}_{s,t}
\end{equation}
to form an eigenvalue problem where $\hat{L}_{s}$ and $\hat{L}_{t}$ are the singlet and triplet dispersion matrices, $\lambda_{s}$ and $\lambda_{t}$ being the eigenvalues corresponding to singlet $\hat{\Phi}_{s}$ and triplet $\hat{\Phi}_{t}$ eigenvectors, respectively.
To find the pair energy, we select $E$, compute $L$ and then $\lambda$. A true pair state corresponds to $\lambda=1$. Thus, all pair energies can be found by adjusting $E$ and searching for $\lambda=1$ using standard binary search algorithms.

As the Green's functions and dispersion matrices become more simplified at the $\Gamma$ point of the FCC lattice, one can take further advantage of this high symmetry point. We use this point to evaluate the binding conditions for the formation of the bound states (with $s$-, $p$-, $d_{T_{2g}}$-, $d_{E_{g}}$- and $f$- symmetries). Via the irreducible representations of the $O_{h}$ group \cite{JFCornwell1997}, we can determine some linear combinations (excluding the normalisation constants) of the eigenvector by performing the 48 operations on the FCC lattice. Note that the eigenfunction $\Phi_{0}^{+}$ is at the centre of zone and therefore remains unchanged due to the operations.
 These operations yield the irreducible representations for both the singlet and triplet states as
\begin{equation}
\begin{split}
& \Gamma_{singlet}^{fcc} = A_{1g} \oplus E_{g} \oplus T_{2g}\\
& \Gamma_{triplet}^{fcc} = T_{1u} \oplus T_{2u} 
\end{split}
\end{equation}
$A_{1g}$ is $s$-symmetrical, $E_{g}$ and $T_{2g}$ are of $d$-symmetry, $T_{1u}$ has $p$-symmetry whilst $T_{2u}$ forms an $f$- symmetric state.
An example of a symmetrised linear combinations for the singlets is 
 
\begin{align}\label{fcc_singlets_linear_combination}
\chi_{}^{A_{1g}} & = \Phi_{1}^{+} + \Phi_{2}^{+} + \Phi_{3}^{+} + \Phi_{4}^{+} + \Phi_{5}^{+} + \Phi_{6}^{+} \\
\chi_{}^{T_{2g}} & = 
\begin{cases}
\Phi_{1}^{+} - \Phi_{4}^{+} \\
\Phi_{3}^{+} - \Phi_{6}^{+} \\
\Phi_{2}^{+} - \Phi_{5}^{+} \\
\end{cases} \\
\chi_{}^{E_{g}} & = 
\begin{cases}
\Phi_{1}^{+} -2\Phi_{2}^{+} + \Phi_{3}^{+} + \Phi_{4}^{+} - 2\Phi_{5}^{+} + \Phi_{6}^{+}\\
\Phi_{1}^{+} -\Phi_{3}^{+} + \Phi_{4}^{+} - \Phi_{6}^{+}
\end{cases}
\end{align}


and for the triplets is
\begin{align}\label{fcc_triplets_linear_combination}
\chi_{}^{T_{1u}} & = 
\begin{cases}
\Phi_{1}^{-} + \Phi_{2}^{-} - \Phi_{4}^{-} + \Phi_{5}^{-} \\
\Phi_{1}^{-} + \Phi_{3}^{-} + \Phi_{4}^{-} - \Phi_{6}^{-}\\
-\Phi_{2}^{-} -\Phi_{3}^{-} + \Phi_{5}^{-} - \Phi_{6}^{-}\\
\end{cases} \\
\chi_{}^{T_{2u}} & = 
\begin{cases}
\Phi_{2}^{-} - \Phi_{3}^{-} - \Phi_{5}^{-} - \Phi_{6}^{-} \\
\Phi_{1}^{-} - \Phi_{3}^{-} + \Phi_{4}^{-} + \Phi_{6}^{-}\\
\Phi_{1}^{-} - \Phi_{2}^{-} - \Phi_{4}^{-} - \Phi_{5}^{-}\\
\end{cases}
\end{align}

Allowing transformation to a new orthogonal basis \footnote{The normalisation factors have been omitted.},\footnote{The subscript $s$ is used twice: $\hat{\Phi}_{s}$ means all possible singlet states (\textsl{s,d,d*}) while $\Phi_{s}$ means an $s$-state only.}
\begin{gather}
\begin{pmatrix}
\Phi_{0}\\ \Phi_{s}\\ \Phi_{d_{1}} \\ \Phi_{d_{2}} \\ \Phi_{d_{3}} \\ \Phi_{d_{4}} \\ \Phi_{d_{5}}
\end{pmatrix}
=
\begin{pmatrix}
1 & 0 & 0 & 0 & 0 & 0 & 0\\
0 & 1 & 1 & 1 & 1 & 1 & 1 \\
0 & 1 & 0 & 0 & -1 & 0 & 0 \\
0 & 0 & 0 & 1 & 0 & 0 & -1 \\
0 & 0 & 1 & 0 & 0 & -1 & 0 \\
0 & 1 & -2 & 1 & 1 & -2 & 1 \\
0 & 1 & 0 & -1 & 1 & 0 & -1
\end{pmatrix}
\begin{pmatrix}
\Phi_{0}^{+}\\
\Phi_{1}^{+}\\
\Phi_{2}^{+}\\
\Phi_{3}^{+}\\
\Phi_{4}^{+}\\
\Phi_{5}^{+}\\
\Phi_{6}^{+}
\end{pmatrix}
\equiv
\hat{\chi_{s}}
\begin{pmatrix}
\Phi_{0}^{+}\\
\Phi_{1}^{+}\\
\Phi_{2}^{+}\\
\Phi_{3}^{+}\\
\Phi_{4}^{+}\\
\Phi_{5}^{+}\\
\Phi_{6}^{+}
\end{pmatrix}
\end{gather}
\begin{gather}
\begin{pmatrix}
\Phi_{p_{1}}\\ \Phi_{p_{2}} \\ \Phi_{p_{3}} \\ \Phi_{f_{1}} \\ \Phi_{f_{2}} \\ \Phi_{f_{3}}
\end{pmatrix}
=
\begin{pmatrix}
1 & 1 & 0 & -1 & 1 & 0 \\
1 & 0 & 1 & 1 & 0 & -1 \\
0 & -1 & -1 & 0 & 1 & -1 \\
0 & 1 & -1 & 0 & -1 & -1 \\
1 & 0 & -1 & 1 & 0 & 1 \\
1 & -1 & 0 & -1 & -1 & 0 
\end{pmatrix}
\begin{pmatrix}
\Phi_{1}^{-}\\
\Phi_{2}^{-}\\
\Phi_{3}^{-}\\
\Phi_{4}^{-}\\
\Phi_{5}^{-}\\
\Phi_{6}^{-}
\end{pmatrix}
\equiv
\hat{\chi_{t}}
\begin{pmatrix}
\Phi_{1}^{-}\\
\Phi_{2}^{-}\\
\Phi_{3}^{-}\\
\Phi_{4}^{-}\\
\Phi_{5}^{-}\\
\Phi_{6}^{-}
\end{pmatrix}
\end{gather}


We diagonalise the equation using
\begin{equation}
\hat{L}_{i}^{diag}=\hat{\chi}_{i}\cdot \hat{L}_{i}\cdot \hat{\chi}_{i}^{-1}
\end{equation}
The respective block-diagonal self-consistent equations are

\begin{align}\label{fcc_diag_dispersion_matrix}
&\colvec[.9]{
UL_{0} & 2VL_{1} & 0 & 0 & 0 & 0 & 0 \\
6UL_{1} & \mathcal{K}_{s} & 0 & 0 & 0 & 0 & 0 \\
0 & 0 & \mathcal{K}_{d_{T_{2g}}} & 0 & 0 & 0 & 0 \\
0 & 0 & 0 & \mathcal{K}_{d_{T_{2g}}} & 0 & 0 & 0 \\
0 & 0 & 0 & 0 & \mathcal{K}_{d_{T_{2g}}} & 0 & 0 \\
0 & 0 & 0 & 0 & 0 & \mathcal{K}_{d_{E_{g}}} & 0 \\
0 & 0 & 0 & 0 & 0 & 0 & \mathcal{K}_{d_{E_{g}}}
}
\colvec[.99]{
\Phi_{0}\\ \Phi_{s}\\ \Phi_{d_{1}}\\ \Phi_{d_{2}}\\ \Phi_{d_{3}} \\ \Phi_{d_{4}}\\ \Phi_{d_{5}}
} = 
\colvec[.99]{
\Phi_{0}\\ \Phi_{s}\\ \Phi_{d_{1}}\\ \Phi_{d_{2}}\\ \Phi_{d_{3}} \\ \Phi_{d_{4}}\\ \Phi_{d_{5}}
}
\end{align}

\begin{gather}\label{fcc_diag_dispersion_matrix2}
\begin{pmatrix}
\mathcal{K}_{p} & 0 & 0 & 0 & 0 & 0 \\
0 & \mathcal{K}_{p} & 0 & 0 & 0 & 0 \\
0 & 0 & \mathcal{K}_{p} & 0 & 0 & 0 \\
0 & 0 & 0 & \mathcal{K}_{f} & 0 & 0 \\
0 & 0 & 0 & 0 & \mathcal{K}_{f} & 0 \\
0 & 0 & 0 & 0 & 0 & \mathcal{K}_{f}
\end{pmatrix}
\begin{pmatrix}
\Phi_{p_{1}}\\ \Phi_{p_{2}} \\ \Phi_{p_{3}} \\ \Phi_{f_{1}} \\ \Phi_{f_{2}} \\ \Phi_{f_{3}}
\end{pmatrix}
=
\begin{pmatrix}
\Phi_{p_{1}}\\ \Phi_{p_{2}} \\ \Phi_{p_{3}} \\ \Phi_{f_{1}} \\ \Phi_{f_{2}} \\ \Phi_{f_{3}}
\end{pmatrix}
\end{gather}
where
\begin{align*}
    \mathcal{K}_{s}&=V[L_{0} + 4L_{1} + L_{2} + 2L_{3} + 4L_{4}]\\
    \mathcal{K}_{d_{T_{2g}}}&=V[L_{0} + L_{2} - 2L_{3}]\\
    \mathcal{K}_{d_{E_{g}}}&=V[L_{0} - 2L_{1} + L_{2} + 2L_{3} - 2L_{4}]\\
    \mathcal{K}_{p}&=V[L_{0} +2L_{1} - L_{2} - 2L_{4}]\\
    \mathcal{K}_{f}&=V[L_{0} -2L_{1} - L_{2} + 2L_{4}]
\end{align*} 

These are the solutions to the two-body problem at $\Pvec=0$. The $2\times2$ block in Eqn. (\ref{fcc_diag_dispersion_matrix}) corresponds to the $s$-symmetrical state, the next three $1\times1$ blocks are triply degenerate $d$- states of $T_{2g}$ symmetry and the last two are another doubly degenerate $d$- states with the $E_g$ symmetry.
In the case of spin triplet states in Eqn. (\ref{fcc_diag_dispersion_matrix2}), the $p$- and $f$- states are 3-fold degenerate and they belong to the $T_{1u}$ and $T_{2u}$ symmetry respectively.\\

It is possible to evaluate the exact binding threshold for the emergence of a bound state. With the symmetrised and diagonalised equations, we set the energy value as $E\rightarrow-2W=-24t$. The self-consistent equations give

\setlength{\belowdisplayskip}{0pt} \setlength{\belowdisplayshortskip}{0pt}
\setlength{\abovedisplayskip}{0pt} \setlength{\abovedisplayshortskip}{0pt}
\begin{align}\label{s_critical_matrix}
s:\hspace{5.8ex}\begin{pmatrix}
1 - UL_{0} & -2VL_1 \\
-6UL_{1} & 1-\mathcal{K}_{s}
\end{pmatrix}=0
\end{align}
\begin{align}\label{critical_v_nonS}
     d_{T_{2g}}:\hspace{13.3ex} &1-\mathcal{K}_{d_{T_{2g}}} =0 \hspace{7ex}\\
     d_{E_{g}}:\hspace{13.3ex} &1-\mathcal{K}_{d_{E_{g}}} =0 \hspace{7ex}\\
     p:\hspace{13.3ex} &1 - \mathcal{K}_{p}= 0 \\
     f:\hspace{13.3ex} &1 - \mathcal{K}_{f}=0 \label{critical_vf}
\end{align}\\

Following Ref. \cite{glasser2000exact}, the exact solution of the Green's functions in Eqn. (\ref{all_gf_at_gamma_point}) are \\
\begin{align}
    L_{0}&=-\frac{\sqrt{3}K_{0}^2}{8\pi^2t} =\frac{-0.056027549298548}{t} \\
    L_{1}&=\frac{1}{24t} -\frac{\sqrt{3}K_{0}^2}{8\pi^2t} =\frac{1}{24t} + L_{0} \\
    L_{2}&=-\frac{9\sqrt{3}K_{0}^2}{8\pi^2t} -\frac{3}{4t\sqrt{3}K_{0}^2} + \frac{2}{3t}= 9L_{0} + \frac{3}{32\pi^2t^2L_{0}} + \frac{2}{3t} \\
    L_{3}&=\frac{\sqrt{3}K_{0}^2}{24\pi^2t}-\frac{1}{8t\sqrt{3}K_{0}^2}=\frac{1}{64\pi^2t^2L_{0}}-\frac{L_0}{3}\\
    L_{4}&=\frac{\sqrt{3}K_{0}^2}{24\pi^2t} +\frac{1}{4t\sqrt{3}K_{0}^2}  -\frac{1}{12t}=-\frac{L_0}{3}-\frac{1}{32\pi^2t^2L_0}-\frac{1}{12t}
\end{align}
where the complete elliptic integral of the first kind $K_{0}=K\left(\frac{\sqrt{3}-1}{2\sqrt{2}}\right)=1.598142002112540$.

The binding conditions are obtained from Eqns. (\ref{s_critical_matrix})--(\ref{critical_vf}) to be

\begin{equation}
    V_{c}^{s}\leq V(U)=\frac{UL_{0}-1}{UL_{0}\mathcal{C}-\mathcal{C}-12UL_{1}^2}
\end{equation}\\
where $\mathcal{C}=L_{0} + 4L_{1} + L_{2} + 2L_{3} + 4L_{4}=12L_0 + \frac{1}{2t}=-0.172330591582576/t$.\\
\begin{align}
    V_c^{d_{T_{2g}}}=-22.734195989010747t\\
    V_c^{d_{E_{g}}}=-26.810644276320041t\\
    V_c^p=-16.302567033831927t \\
    V_c^f=-27.416574191996979t
\end{align}
In specific limits, the critical binding of $s$-states is
\begin{align}
    &V_{c}^{s}(U=0) =-5.80280025t\\
    &V_{c}^{s}(U\rightarrow+\infty)=-7.8028002504t \\
    &U_{c}(V=0) =-17.84836232388t \\
    &U_{c}(V\rightarrow+\infty) =-24t
\end{align}

%

\section{Pair mass in the superlight limit}\label{appendix:superlightmass}

Expanding the one-particle dispersion at small ${\kvec}$, one obtains 
\begin{equation}
\varepsilon_{\kvec} \approx - 12t + t b^2 ( k^2_x + k^2_y + k^2_z ) = 
\varepsilon_0 + \frac{\hbar^2}{2 m_0} \, ( k^2_x + k^2_y + k^2_z ) \, ,  
\label{twopart:eq:appgthree}
\end{equation}
where $m_0 = \hbar^2/(2tb^2)$ is the free particle mass. 

We define six singlet dimer basis states 
\begin{equation}
D_{i,{\nvec}} = \frac{1}{\sqrt{2}} \left( 
                \left\vert   \uparrow \right\rangle_{{\nvec}}
                \left\vert \downarrow \right\rangle_{{\nvec} + {\avec}_i}  + 
                \left\vert \downarrow \right\rangle_{{\nvec}}
                \left\vert   \uparrow \right\rangle_{{\nvec} + {\avec}_i} \right) . 
\label{twopart:eq:appgfour}
\end{equation}
$D_{i,{\nvec}}$ are the only states with nonzero weights in the $V \rightarrow -\infty$ limit. Because of the topology of the FCC lattice, $D_{i,{\nvec}}$ are linked by first-order hopping events. The first-order Hamiltonian matrix is
\begin{eqnarray}
\hat{H} D_{1,{\nvec}} & = & - t \left( D_{3,{\nvec}} + D_{3,{\nvec}+{\avec}_6} \right) 
                            - t \left( D_{4,{\nvec}} + D_{4,{\nvec}+{\avec}_5} \right)
                            - t \left( D_{5,{\nvec}} + D_{5,{\nvec}+{\avec}_4} \right)    
                            - t \left( D_{6,{\nvec}} + D_{6,{\nvec}+{\avec}_3} \right)                            
\label{twopart:eq:appgfive}  \\
\hat{H} D_{2,{\nvec}} & = & - t \left( D_{3,{\nvec}} + D_{3,{\nvec}-{\avec}_5} \right) 
                            - t \left( D_{4,{\nvec}} + D_{4,{\nvec}-{\avec}_6} \right)
                            - t \left( D_{5,{\nvec}+{\avec}_2} + D_{5,{\nvec}-{\avec}_5} \right)    
                            - t \left( D_{6,{\nvec}+{\avec}_2} + D_{6,{\nvec}-{\avec}_6} \right)   
\label{twopart:eq:appgsix}   \\
\hat{H} D_{3,{\nvec}} & = & - t \left( D_{1,{\nvec}} + D_{1,{\nvec}-{\avec}_6} \right) 
                            - t \left( D_{2,{\nvec}} + D_{2,{\nvec}+{\avec}_5} \right)
                            - t \left( D_{5,{\nvec}} + D_{5,{\nvec}+{\avec}_2} \right)    
                            - t \left( D_{6,{\nvec}+{\avec}_3} + D_{6,{\nvec}-{\avec}_6} \right)   
\label{twopart:eq:appgseven} \\
\hat{H} D_{4,{\nvec}} & = & - t \left( D_{1,{\nvec}} + D_{1,{\nvec}+{\avec}_2} \right) 
                            - t \left( D_{2,{\nvec}} + D_{2,{\nvec}+{\avec}_6} \right)
                            - t \left( D_{5,{\nvec}-{\avec}_5} + D_{5,{\nvec}+{\avec}_4} \right)    
                            - t \left( D_{6,{\nvec}} + D_{6,{\nvec}+{\avec}_2} \right)   
\label{twopart:eq:appgeight} \\
\hat{H} D_{5,{\nvec}} & = & - t \left( D_{1,{\nvec}} + D_{1,{\nvec}-{\avec}_4} \right) 
                            - t \left( D_{2,{\nvec}-{\avec}_2} + D_{2,{\nvec}+{\avec}_5} \right)
                            - t \left( D_{3,{\nvec}} + D_{3,{\nvec}-{\avec}_2} \right)    
                            - t \left( D_{4,{\nvec}-{\avec}_4} + D_{4,{\nvec}+{\avec}_5} \right)   
\label{twopart:eq:appgnine} \\
\hat{H} D_{6,{\nvec}} & = & - t \left( D_{1,{\nvec}} + D_{1,{\nvec}-{\avec}_3} \right) 
                            - t \left( D_{2,{\nvec}-{\avec}_2} + D_{2,{\nvec}+{\avec}_6} \right)
                            - t \left( D_{3,{\nvec}+{\avec}_6} + D_{3,{\nvec}-{\avec}_3} \right)    
                            - t \left( D_{4,{\nvec}} + D_{4,{\nvec}-{\avec}_2} \right)   
\label{twopart:eq:appgten}
\end{eqnarray}
Applying a Fourier transform one obtains the dimer Schr\"odinger equation. Its self-consistent condition yields pair energy $E$ for a given pair momentum $\Pvec$. 
\begin{equation}
\left\vert \begin{array}{cccccc}
E   &   0   &   t ( 1 + e^{i\Pvec{\avec}_6} )   &  t ( 1 + e^{i\Pvec{\avec}_5} )  
            &   t ( 1 + e^{i\Pvec{\avec}_4} )   &  t ( 1 + e^{i\Pvec{\avec}_3} ) \\
0   &   E   &   t ( 1 + e^{-i\Pvec{\avec}_5} )  &  t ( 1 + e^{-i\Pvec{\avec}_6} )  
            &   t ( e^{i\Pvec{\avec}_2} + e^{-i\Pvec{\avec}_5} )  
            &   t ( e^{i\Pvec{\avec}_2} + e^{-i\Pvec{\avec}_6} )                 \\
t ( 1 + e^{-i\Pvec{\avec}_6} )  &   t ( 1 + e^{i\Pvec{\avec}_5} )  &  E  &  0 
            &   t ( 1 + e^{i\Pvec{\avec}_2} )   
            &   t ( e^{i\Pvec{\avec}_3} + e^{-i\Pvec{\avec}_6} )                 \\            
t ( 1 + e^{-i\Pvec{\avec}_5} )  &   t ( 1 + e^{i\Pvec{\avec}_6} )  &  0  &  E 
            &   t ( e^{i\Pvec{\avec}_4} + e^{-i\Pvec{\avec}_5} )   
            &   t ( 1 + e^{i\Pvec{\avec}_2} )                                      \\
t ( 1 + e^{-i\Pvec{\avec}_4} )  &  t ( e^{i\Pvec{\avec}_5} + e^{-i\Pvec{\avec}_2} )
            &  t ( 1 + e^{-i\Pvec{\avec}_2} )  
            &  t ( e^{i\Pvec{\avec}_5} + e^{-i\Pvec{\avec}_4} )   &  E  &  0     \\
t ( 1 + e^{-i\Pvec{\avec}_3} )  &  t ( e^{i\Pvec{\avec}_6} + e^{-i\Pvec{\avec}_2} ) 
            &  t ( e^{i\Pvec{\avec}_6} + e^{-i\Pvec{\avec}_3} )
            &  t ( 1 + e^{-i\Pvec{\avec}_2} )    &   0   &   E 
\end{array} \right\vert = 0  
\label{twopart:eq:appgeleven}
\end{equation}

%
The general dispersion Eqn.~(\ref{twopart:eq:appgeleven}) is too complex. However, to extract a pair mass it is sufficient to know $E(\Pvec)$ at small $\Pvec$. Utilising the isotropy property of cubic dispersion relations, we set $\Pvec = ( P_x , 0 , 0 )$ and expand Eqn.~(\ref{twopart:eq:appgeleven}) to get
\begin{equation}
 E^3 ( E - 4t ) \left[ E^2 + 4 t E - 16 t^2 \left( 1 + \cos{\frac{P_x b}{2}} \right) \right] = 0 \: ,
\label{twopart:eq:appgtwelve}
\end{equation}
which defines dispersion of six pair bands along the $P_x$ direction. The lowest band is 
\begin{equation}
 E_1 (P_x) = - 2t \left( 1 + \sqrt{ 5 + 4 \cos{\frac{P_x b}{2}} } \right) .
\label{twopart:eq:appgthirteen}
\end{equation}
Expanding at small $P_x$, one obtains
\begin{equation}
 E_1 ( P_x b \ll 1 ) \approx - 8t + \frac{1}{6} \, t (P_x b)^2 
 \equiv E_0 + \frac{\hbar^2 P^2_x}{2 m^{\ast}} \: ,
\label{twopart:eq:appgfourteen}
\end{equation}
from where
\begin{equation}
m^{\ast} = \frac{3 \hbar^2}{t b^2} = 6 m_0 \: .
\label{twopart:eq:appgfifteen}
\end{equation}
Thus even an infinitely bound intersite pair is only six times heavier than a free particle.


\end{document}